\documentclass[aps,pra,groupedaddress,amsmath,amssymb,twocolumn]{revtex4-1}

\usepackage{amsmath}
\usepackage[utf8]{inputenc}
\usepackage[T1]{fontenc}
\usepackage[english]{babel}
\usepackage[stable]{footmisc}
\usepackage[english]{hyperref}
\usepackage{multirow}
\usepackage{booktabs}
\usepackage{graphicx}
\usepackage{float}
\usepackage{units}
\usepackage{upgreek}
\usepackage{mathtools}
\usepackage{subfigure}
\usepackage{paralist}
\usepackage{xcolor}
\usepackage{amssymb}

\usepackage[a4paper, total={16cm, 27cm}]{geometry}

\usepackage{todonotes}

\begin{document}

\title{Predicting properties of the stationary probability currents for two-species reaction systems without solving the Fokker-Planck equation}

\author{Marc Mendler, Barbara Drossel}
\date{\today}

\affiliation{Institut für Festkörperphysik, Technische Universität Darmstadt, Hochschulstr. 6, 64289 Darmstadt, Germany}

\begin{abstract}
We derive methods for estimating the topology of the stationary probability current $\vec{j}_s$ of the two-species Fokker-Planck equation (FPE) without the need to solve the FPE. These methods are chosen such that they become exact in certain limits, such as infinite system size or vanishing coupling between species in the diffusion matrix. The methods make predictions about the fixed points of $\vec{j}_s$ and their relation to extrema of the stationary probability distribution and to fixed points of the convective field,  which is related to the deterministic drift of the system. Furthermore, they predict the rotation sense of $\vec{j}_s$ around extrema of the stationary probability distribution. Even though these methods cannot be proven to be valid away from extrema, the boundary lines between regions with different rotation sense are obtained with surprising accuracy. We illustrate and test these method using simple reaction systems with only one coupling term between the two species as well as a few generic reaction networks taken from literature. We use it also to investigate the shape of non-physical probability currents occurring in reaction systems with detailed balance due to the approximations involved in deriving the Fokker-Planck equation.

\end{abstract}


\maketitle

\section{Introduction}
Reaction networks of interacting species are used as models in chemistry, synthetic biology, epidemiology, ecology, and even sociology.

If the investigated system is large enough and stochastic effects can be neglected, the dynamics of such networks can be described by a set of ordinary differential equations. The properties of these models can be investigated with methods from the theory of dynamical systems \cite{strogatz2014nonlinear}. However, these models become insufficient if the number of particles (or individuals) of the involved species becomes small. Then the stochastic change of particle numbers acts as a source of intrinsic noise that can modify the behaviour of the deterministic model \cite{mckane2014ants, plesa2017noise, boland2008limit}. In order to account for such effects, a stochastic modeling approach is needed.

\smallskip

One very general mathematical description of such stochastic reaction networks is the so-called master equation. The master equation is a continuous-time Markov model of the system and yields a time-dependent probability distribution for the possible system states \cite{kampen, gillespie_buch}. While the master equation can be simulated on a computer using stochastic simulation algorithms \cite{gillespie_algorithm}, solving it analytically is impossible for most systems. If one does not want to rely solely on computer simulations, it is necessary to use approximation methods.
\smallskip

One particularly well-known approximation of the master equation is the Fokker-Planck equation (FPE). The FPE is derived through the Kramers-Moyal expansion, where the integer particle numbers are replaced by continuous substrate concentrations and an expansion for small step sizes is carried out. The first two terms of this expansion, named deterministic drift and diffusion, give the FPE \cite{gardiner}. Interestingly, the solution of the FPE is a valid probability density, while for all higher order truncations of the Kramers-Moyal expansion non-physical probability densities can occur that can become negative or are not normalizable \cite{risken1996fokker}.

\smallskip

Even though the assumptions used for deriving the FPE becomes invalid for small particle concentrations, the FPE can still describe many of such small systems quite accurately in cases where the corresponding deterministic models already break down \cite{grima2011accurate, alonso2006stochastic, mckane2014ants, mckane2004stochastic, falk2017schloegl}. 
As the FPE is furthermore much easier to solve than the master equation, it is a very useful tool to describe stochastic effects such as shifts of fixed points, stochastic cycles, mean first passage times, and stochastic $p$-bifurcations  \cite{mckane2014ants, mckane2005resonance, gardiner, erban_analysis_2008, falk2017schloegl}. Furthermore, it can be utilized to derive stochastic phase portraits that give information about the extrema of the stationary probability distribution \cite{mendler2018analysis}.

Despite these successes, there are several phenomena that the FPE does not describe correctly. These include among others a certain type of noise-induced multistability in gene regulation networks with slow switching between the ON and OFF states \cite{duncan2015noise}, some phenomenological bifurcations induced by reducing the particle number \cite{becker2020relation}, or other phenomena where very small particle numbers are involved \cite{kampen, grima2011accurate, gillespie2000chemical}.

One particularly interesting flaw of the FPE is 
that there are systems for which detailed balance holds for the master equation, while the FPE shows non-vanishing probability currents in its steady state \cite{ceccato2018remarks}. Such probability currents are non-physical as they are an artifact caused by the approximations involved in the FPE.  The properties of such non-physical probability currents, such as their topology and the circumstances under which they emerge, have not yet been fully investigated.  

\smallskip

Even more importantly, stationary probability currents in general have attracted an increasing attention of researchers in recent years. The reason is that most reaction systems are driven systems that approach a non-equilibrium steady state with a non-vanishing energy throughput. Such systems include metabolic networks, food webs, or economic models. 

In order to calculate the stationary probability currents that characterize the long-term behaviour of these systems, it is in general necessary to solve the Fokker-Planck equation numerically. This makes these currents much less accessible for a quick qualitative analysis than e.g.~the extrema of the probability density, which can be calculated analytically for 1D systems and can often be obtained approximately for 2D systems \cite{gillespie_buch, falk2017schloegl, mendler2018analysis}.

It is the goal of the present paper to provide tools for estimating the topology of the stationary probability current $\vec{j}_s$ directly from the deterministic drift and diffusion matrix of the system without the need to solve the FPE. An important quantity in our calculations is the convective field, which is the sum of the deterministic drift and systematic contributions to the current due to the diffusion term. 

We will proceed in several steps. First, we will derive an explicit expression for $\vec{j}_s$ and the stationary probability distribution that is valid to leading order in the inverse system size $1/N$. This will allow us to obtain a relation between the fixed points of $\vec{j}_s$, those of the convective field, and the extrema of the stationary probability distribution, which is again valid to leading order in $1/N$. By considering artificial FPEs that contain only one or few of the possible interaction terms, we will demonstrate the usefulness of these methods and will show that when the leading order terms in $1/N$ vanish, dipole currents will emerge. We will also discuss simple realistic reaction systems, in particular one that shows a limit cycle. 

Finally we will utilize our method to investigate the mentioned non-physical stationary probability currents resulting from the FPE. In one example, we will find stationary currents that take the shape of a  quadrupole. 

During all our investigations, we will compare the results of our method to the actual topology of $\vec{j}_s$ obtained from the numerical solution of the corresponding FPE.

\section{Methods}
\label{sec:methods}

\subsection{Chemical reactions with mass action kinetics}
We consider $M$ species $X_i$ that undergo a set of $R$ (chemical) reactions labeled by~$r$,
\begin{align}
\label{eq:reaction_system}
\sum\limits_{i=1}^M \sigma_{ir} X_i &\overset{k_r}{\longrightarrow} \sum\limits_{j=1}^M \tau_{jr} X_j
\end{align}
with stoichiometric constants $\sigma_{ir}$ and $\tau_{ir}$ and reaction constants $k_r$.

We denote the number of individuals (or molecules) of species $X_i$ as $n_i$. The state of the system is completely determined by the state vector $\vec{n}$ which gives the number of molecules of each species. To compactify notation, one defines the stoichiometric matrix \cite{kampen}
\begin{equation}
	S_{ir} = \tau_{ir} - \sigma_{ir}
\end{equation}
 and the propensity vector
\begin{align}
	\nu_r(\vec{n}) &= k_r \prod_{i=1}^M  N^{-\sigma_{ir}}  \frac{n_i!}{(n_i-\sigma_{ir})!}\nonumber\\
	&\approx k_r \prod_{i=1}^M  \left(\frac{n_i}{N}\right)^{\sigma_{ir}} \text{for } n_i \gg \sigma_{ir}\, .
	\label{eq:nu}
\end{align}
Here, $N$ denotes the system size, which equals the total number of particles inside the system, in reaction systems where this number is preserved.
In systems where the total number of particles can change, $N$ can be interpreted as a dimensionless quantity proportional to the reactor volume \cite{kampen}.

Using this quantity we can define $x_i := \frac{n_i}{N}$ as the concentration of species $X_i$. As the propensity vector $\nu_r(\vec{n})$ contains the product of these concentrations, it is proportional to the probability that the right molecules for reaction $r$ meet at the same time in the same small volume element. 

\subsection{Chemical Master Equation}
The master equation yields a stochastic description of the time evolution of the state vectors $\vec{n}$ of such chemical reactions.
It describes the probability flow between the different system states and thus the change of probability $P(\vec{n}, t)$ of theses states,
\begin{equation}
    \frac{\text d P(\vec{n}, t)}{\text d t} = \sum\limits_r \left[ \nu_r (\vec{n} - \vec{S}_r) P(\vec{n} - \vec{S}_r, t) - \nu_r(\vec{n})P(\vec{n}, t)\right]
    \label{eq:me}
\end{equation}
The sum over all reactions can be transformed into a sum over all system states,
\begin{equation}
    \frac{\text d P(\vec{n}, t)}{\text d t} = \underbrace{\sum\limits_{\vec{m}} \mu (\vec{n}|\vec{m}) P(\vec{m}, t)}_{J_{in}} - \underbrace{\sum\limits_{\vec{m}} \mu( \vec{m}|\vec{n})P(\vec{n}, t)}_{J_{out}}\, .
    \label{eq:me2}
\end{equation}
This expression contains the transition rates
\begin{equation}
\mu(\vec{n}|\vec{m})=\sum\limits_{r:\; \vec{n}-\vec{m} = \vec{S}_r}\nu_r(\vec{n}-\vec{S}_r)\, ,
\label{eq:me:defMu}
\end{equation}
and is a more general form of the master equation since it is not confined to systems where the transitions between states occur via chemical reactions.
The probability to be in state $\vec{n}$ changes due to the inflow $J_{in}$ of probability from neighboring states minus the outflow $J_{out}$ of probability from  state $\vec{n}$.
\smallskip

A steady state $P_s(\vec{n})$  is reached whenever $J_{in}$ and $J_{out}$ cancel  at every point $\vec{n}$ in phase space. If additionally not only the sums in \eqref{eq:me2} are equal, but if the terms for each value of $\vec m$ cancel separately, i.e., if
\begin{equation}
    \mu(\vec{n}|\vec{m}) P(\vec{m},t) = \mu(\vec{m}|\vec{n}) P(\vec{n},t) \, ,
\end{equation}
the steady state satisfies detailed balance. 

\subsection{Fokker-Planck equation}
\label{sec:multi_FPE}
A well-known approximation method for the master equation is the Fokker-Planck equation (FPE), where the discrete state space of the reaction system is approximated as continuous. This approximation becomes more accurate when the system size becomes larger.
\smallskip

The FPE is derived from the master equation by performing the so called Kramers-Moyal expansion \cite{gardiner}. To simplify the notation, one introduces the deterministic drift $\vec{f} = \mathbf{S}\cdot \vec{\nu}$ and the diffusion matrix $\mathbf{D} = \mathbf{S}\cdot\text{diag}(\vec{\nu})\cdot\mathbf{S}^T$. With these definitions the FPE reads
\begin{align}
\begin{aligned}
\frac{\partial p(\vec{x},t)}{\partial t} &= - \sum_i \frac{\partial}{\partial x_i}\left[f_i(\vec{x})p(\vec{x},t)\right]\\
&+\frac{1}{2 N}\sum_{ij}\frac{\partial^2}{\partial x_i \partial x_j}\left[D_{ij}(\vec{x})p(\vec{x},t)\right]\, ,
\label{eq:fpe}
\end{aligned}
\end{align}
where $x_i$ denotes again the concentration of species $X_i$.
Here $p$ is a probability density, in contrast to the probability $P$ in the master equation.

The FPE can also be written in the form of a continuity equation, $\frac{\partial p}{\partial t} = - \vec{\nabla} \cdot \vec{j}$ with the probability current 
\begin{equation}
\vec{j} = \vec{\alpha}(\vec{x}) p(\vec{x},t) - \frac{1}{2 N}\mathbf{D}\cdot\vec{\nabla}p(\vec{x},t)
\label{eq:fpe:j_definition}
\end{equation}
and the convective field $\vec{\alpha}(\vec{x})$
\begin{equation}
\vec{\alpha}(\vec{x}) = \vec{f}(\vec{x}) - \frac{1}{2 N} \sum\limits_{ik}  \frac{\partial D_{ik}}{\partial x_k} \vec{\epsilon_i} \, ,
\label{eq:alpha}
\end{equation}
where $\vec{\epsilon}_i = \vec{x}_i/|x_i|$ denotes the unit vector in direction $x_i$  \cite{mendler2018analysis}.
In this notation it is obvious that the system size $N$ determines the strength of the intrinsic noise: with increasing $N$, the diffusion (noise) term decreases. For $N \rightarrow \infty$ the diffusion term vanishes and the system becomes deterministic.

The stationary distribution of the FPE is calculated via $\frac{\partial p_s(\vec{x},t)}{\partial t}=0$, which means $\vec{\nabla} \cdot \vec{j}_s=0$. In a one-dimensional closed system, where $\vec j$ must vanish at the boundaries, this implies $\vec j_s=0$. In higher-dimensional systems, stationary states can have $\vec j_s \neq 0$, as the condition  $\vec{\nabla} \cdot \vec{j}_s=0$ can be satisfied with nonzero currents. Such stationary solutions constitute non-equilibrium steady states. For these states the steady-state condition $\vec{\nabla} \cdot \vec{j}_s=0$ implies that $\vec j_s = \vec \nabla \times \vec A$ with a suitable vector field $\vec A(\vec x)$, i.e., $\vec{j}_s$ must be solenoidal.

\subsection{Condition for $\vec{j}_s = 0$}
\label{sec:int_cond}

Calculating the stationary probability current $\vec{j}_s$ for a given reaction system is a computationally costly task, as it requires solving the stationary Fokker-Planck equation. One can however easily determine whether the stationary state is an equilibrium state (i.e., $\vec{j}_s=0$): Applying the equilibrium condition to definition \eqref{eq:fpe:j_definition}, we obtain
\begin{align}
    0 = \vec{j_s} &= \vec{\alpha} p_s - \frac{1}{2 N} \mathbf{D}\vec{\nabla} p_s \label{eq:fpe:j:ss}\, .
\end{align}
This can be transformed to $ \mathbf{D}^{-1} \vec{\alpha} = \vec{\nabla}\phi_s $ with $\phi = \frac{1}{2 N} \ln(p)$, which in turn implies $\vec{\nabla}\times\mathbf{D}^{-1}\vec{\alpha} = 0$.
Therefore we arrive at  \cite{gardiner}
\begin{equation}
    \vec{j}_s=0 \phantom{mm} \Leftrightarrow \phantom{mm} \vec{\nabla} \times \mathbf{D}^{-1}\vec{\alpha} = 0
    \label{eq:int_cond}
\end{equation}
The reverse direction is obtained via explicit construction of $p_s$: From $\vec{\nabla}\times\mathbf{D}^{-1}\vec{\alpha} = 0$ we obtain $\mathbf{D}^{-1} \vec{\alpha} = \vec{\nabla}\phi$ with a scalar function $\phi$ that is unique up to an additive constant. By setting $p_s=Ce^{\phi}$ with the normalization constant $C$ that fixes unequivocally the additive constant of $\phi$, we have found the stationary solution of  equation \eqref{eq:fpe:j:ss}.

\bigskip

Relation \eqref{eq:int_cond} is often called the integrability condition.
Besides this relation, there is one simple case where we can immediately see that $\vec{j}_s$ vanishes: If the species are uncoupled, we effectively deal with multiple one-dimensional systems, which always fulfill $\vec{j}_s = 0$.

It should be mentioned that $\vec{j}_s = 0$ is often associated with detailed balance in the master equation \cite{gardiner}. However, the stationary current $j_s$ obtained from the Fokker-Planck equation is only an approximation to the full stationary current, as it is obtained by a second order truncation of the Kramers-Moyal expansion. As Ceccato et al. have shown \cite{ceccato2018remarks}, there are systems that show detailed balance for the stationary master equation while $\vec{j}_s \neq 0$ in the Fokker-Planck description. Conversely it is possible to construct reaction networks with $\vec{j}_s = 0$ but without detailed balance, as for example the reaction system
\begin{align}
    \begin{aligned}
    \emptyset &\rightarrow X\\
    2 X &\rightarrow \emptyset
    \end{aligned}
\end{align}
where $\vec{j}=0$ must hold since the system is one-dimensional while detailed balance can be ruled out using Kolmogorov's criterion \cite{kelly1979reversibility}.

\subsection{Deterministic rate equations and linear stability analysis}
\label{sec:lin_stab}

For very large systems, it often sufficient to use the deterministic description given by the rate equation
\begin{equation}
    \frac{\text d \vec{x}}{\text d t} = \vec{f}(\vec{x})\, ,
    \label{eq:dynamic_system}
\end{equation}
with the deterministic drift $\vec{f}(\vec{x})$  defined in section \eqref{sec:multi_FPE}.

This ordinary differential equation is much easier to solve than the Fokker-Planck equation. A very important technique to analyze \eqref{eq:dynamic_system} is linear stability analysis, where the system is linearized around the fixed points of $\vec{f}$, in order to evaluate their stability.

Linearizing \eqref{eq:dynamic_system} around a fixed point $\vec{x}_0$ yields
\begin{equation}
    \frac{\text d \vec{x}}{\text d t} = J_{\vec{f}}(\vec{x}_0) \cdot (\vec{x} - \vec{x}_0) + \mathcal{O}(\vec{x} - \vec{x}_0)^2
\end{equation}
with the Jacobian $J_{\vec{f}}^{ij}(\vec{x}_0) = \frac{\partial f_i}{\partial x_j} (\vec{x}_0)$. The general solution of this linearized equation is a linear superposition of exponential functions, with the coefficients in the exponent being the eigenvalues of $J_{\vec{f}}(\vec{x}_0)$.  If the real part of all eigenvalues is negative, the fixed point is stable. If at least one  eigenvalue have a positive real part, the fixed point is unstable. 

In two dimensions the eigenvalues are given by
\begin{equation}
    \lambda_{1/2} = \frac{1}{2}\left( \tau \pm \sqrt{\tau^2 - 4 \Delta}\right)\label{eq:eigenvalues1}
\end{equation}
with $\tau = \text{tr}(J_{\vec{f}}) = \vec{\nabla}\cdot \vec{f} = \frac{\partial f_1}{\partial x} + \frac{\partial f_2}{\partial y}$ and $\Delta = \text{det}(J_{\vec{f}}) = \frac{\partial f_1}{\partial x} \frac{\partial f_2}{\partial y} - \frac{\partial f_2}{\partial x} \frac{\partial f_1}{\partial y}$. 
Complex eigenvalues occur if $\tau^2 < 4\Delta$ and are associated with spirals or centers. They can occur only if the derivatives $\frac{\partial f_2}{\partial x}$ and $\frac{\partial f_1}{\partial y}$ have opposite signs, which means in particular that  $\vec{\nabla}\times\vec{f} = \frac{\partial f_2}{\partial x} - \frac{\partial f_1}{\partial y}$ must be non-zero.
Later, we will make use of this insight that a fixed point of $\vec{f}$ can only be a spiral or center if $\vec{\nabla}\times\vec{f} \neq 0$.

Another important consequence of linear stability analysis is that the fixed points of  a solenoidal vector field $\vec{j}$ with $\vec{\nabla}\cdot \vec{j} = 0$ must be saddle points or centers: For $\tau = \vec{\nabla}\cdot\vec{j} = 0$, the eigenvalues are
\begin{equation}
    \lambda_{1/2} = \pm \sqrt{-\Delta}\, ,
\end{equation}
which can either be two real numbers of opposite sign  for $\Delta > 0$, indicating a saddle point, or two complex conjugate numbers with vanishing real part, indicating a center.

\section{Results}

In the following, we will derive several analytical results concerning the properties of the stationary current $\vec{j}_s$, supplemented by a numerical analysis of generic cases and specific models.

\subsection{Approximating $\vec{j}_s$ for large $N$}
\label{sec:leading_order_n}

For large system sizes the convective field $\vec\alpha$ deviates only little from the deterministic drift $\vec f$, and the probability current $\vec{j}$ deviates only little from the deterministic case, where all current flows to the attractors of $\vec f$, which are either stable fixed points or limit cycles. This means in particular that for sufficiently large system size $N$ fixed points of $\vec j_s$ and those of $\vec \alpha$ must be closely related. We perform in the following an expansion to leading order in $1/N$ in order to obtain an analytical expression for the stationary current in terms of $\vec \alpha$ and $\mathbf{D}$. 

To this purpose, we split the convective field into two contributions 
\begin{equation}\label{eq:alphazerlegung}
\vec{\alpha}= \mathbf{D} \cdot \vec{\alpha}_{\parallel} + \vec{\alpha}_{\perp}
\end{equation}
with
\begin{align}
     \vec{\alpha}_ {\perp} \perp \vec{\alpha}_{\parallel}\label{eq:alpha_perp_condition}
\end{align}
and
\begin{equation}
    \vec{\nabla}\times\vec{\alpha}_{\parallel}
    \label{eq:def:alpha_parallel} = 0\, .
\end{equation}

In the case that $\mathbf{D} $ is proportional to the identity matrix, the calculation of  $\vec{\alpha}_{\parallel}$ and $\vec{\alpha}_{\perp}$ can be done straightforwardly in Fourier space by requiring that the conditions \eqref{eq:alpha_perp_condition}  and \eqref{eq:def:alpha_parallel} are satisfied for each Fourier component separately, which gives 
\begin{equation}
    \vec k \times \vec{\alpha}_{\parallel,\vec k} =0\, .
\end{equation}
This fixes the plane in which $\vec{\alpha}_{\parallel_{\vec k}}$ must lie, and this in turn fixes the decomposition into the two perpendicular vectors $\vec{\alpha}_{\vec k} = \vec\alpha_{\perp,\vec k} +  \vec{\alpha}_{\parallel,\vec k}$. 

\smallskip

We will now use the decomposition \eqref{eq:alphazerlegung} to find explicit expressions for $p_s$ and $\vec j_s$.  Let us first consider the special case that  $\vec{\alpha}_{\perp} = 0$. Then the relations  \eqref{eq:def:alpha_parallel} and \eqref{eq:int_cond} yield that $\vec{j}_s$ vanishes. The definition  \eqref{eq:fpe:j_definition} then gives immediately the stationary probability density
\begin{equation}
    p_s \equiv p_s^{\parallel} = A e^{2 N \int \vec{\alpha}_{\parallel} \text d \vec{x}}
    \label{eq:analytisch:p_solution}
\end{equation}
and thus
\begin{equation}
    \vec{\nabla} p_s^{\parallel} = 2 N \vec{\alpha}_{\parallel} \cdot p_s^{\parallel}\, .
    \label{eq:analytisch:p_grad}
\end{equation}
Not surprisingly, the stationary distribution becomes narrower and higher with increasing $N$, and the ratio of $\vec\nabla p$ and $p$ is proportional to $N$.
The $N$ dependence of the two contributions to $\vec{j}$ is therefore of the same order,
\begin{equation}
    \vec{j} = \underbrace{\vec{\alpha}p}_{\thicksim p} - \underbrace{\frac{1}{2 N} \mathbf{D} \vec{\nabla} p}_{\thicksim p}\, ,
    \label{eq:j_definition_with_orders}
\end{equation}
and the terms occurring in its divergence differ in their $N$ dependence:
\begin{equation}\label{eq:divj_1N}
    0 \overset{!}{=} \vec{\nabla}\cdot\vec{j} = \underbrace{\vec{\nabla}\vec{\alpha} \cdot p}_{\thicksim p} + \underbrace{\vec{\alpha} \cdot \vec{\nabla} p}_{\thicksim Np} - \underbrace{\frac{1}{2 N} \vec{\nabla} (\mathbf{D} \vec{\nabla} p)}_{\thicksim Np} .
\end{equation}
This means that in leading order we can neglect the first term. Having dropped the first term, we  now insert the partition of $\vec{\alpha}$ defined above, obtaining
\begin{equation}
    0 = \vec{\alpha}_{\perp} \cdot \vec{\nabla} p + \mathbf{D} \vec{\alpha}_{\parallel} \cdot \vec{\nabla} p + \frac{1}{2 N} \vec{\nabla} (\mathbf{D} \vec{\nabla} p)\, .
\end{equation}

The solution of this partial differential equation is to leading order still given by \eqref{eq:analytisch:p_solution}:
\begin{equation}
    0 = 2 N p_s^{\parallel} \cdot \underbrace{\vec{\alpha}_{\perp} \cdot\vec{\alpha}_{\parallel}}_0  + \vec{\nabla} \underbrace{(\vec{\alpha}_{\parallel} p_s^{\parallel} - \frac{1}{2N}\mathbf{D}\vec{\nabla}p_s^{\parallel})}_0 .
\end{equation}

This means that to leading order in $1/N$, $\vec{\alpha}_{\perp}$ has no influence on $p_s$. Thus we can insert \eqref{eq:alphazerlegung} and \eqref{eq:analytisch:p_grad} in \eqref{eq:j_definition_with_orders} and obtain
\begin{equation}
    \vec{j}_s = \vec{\alpha}_{\perp} p_s
    \label{eq:j:leading_order}
\end{equation} in leading order in $1/N$.
This has interesting implications:

First, due to \eqref{eq:alpha_perp_condition} and \eqref{eq:analytisch:p_grad} $\vec{j}_s$ is always perpendicular to $\vec{\nabla}p_s$ and thus follows the height lines of $p_s$. Consequently, $\vec{j}_s$ must have  a fixed point at the location of maxima and minima of $p_s$ if it does not vanish. Conversely, all maxima and minima of $p_s$ will lie at the location of $\vec{j_s}$ fixed points. The same is of course also true for saddle points of $p_s$ and $\vec{j}_s$ in this leading order approximation in $1/N$. 
\smallskip

Because of eq.~\eqref{eq:fpe:j_definition}, the correspondence between $\vec{j}_s$ fixed points and $p_s$ extrema means that $\vec{\alpha}$ has fixed points at their locations, too. Therefore we obtain for large $N$ the relation
\begin{equation}
    \vec{j}_s (\vec{x}_0) = 0 \Leftrightarrow \vec{\alpha}(\vec{x}_0) = 0 \, .
    \label{eq:j_alpha_correspondence}
\end{equation}

Taken together, we have derived the three relations \eqref{eq:analytisch:p_solution}, \eqref{eq:j:leading_order}, and \eqref{eq:j_alpha_correspondence}, which are valid to leading order in $1/N$, which means that they are a good approximation for sufficiently large system sizes. For smaller system sizes, or when the leading-order term vanishes, higher-order terms become important. 

In the following, we will first explore further consequences of this approximation. Then, we will discuss  phenomena that can not be explained based on this first-order approximation, such as  dipole currents in $\vec{j}_s$, slow transient states where $p_s$ has a maximum but $\vec{j}_s$ does not vanish, and saddle points in $p_s$ where $\vec{j}_s$ cannot form a saddle point due to topological constraints.

\subsection{Correspondence between $\vec{\alpha}$ and $\vec{j}$ fixed points for large $N$}
\label{sec:fp_correspondence}

Next, we will identify the type of $\vec{j}_s$ fixed point associated with each type of $\vec{\alpha}$ fixed point, based on the leading order in $1/N$. We have already seen that in this approximation $\vec{j}_s$ follows the height lines of $p_s$. This means that $\vec{j}_s$ follows closed loops around maxima and minima of $p_s$, and consequently the fixed  point of $\vec{j}_s$ at these extrema of $p_s$ is a center. On the other hand, maxima of $p_s$ are associated with stable fixed points of $\vec\alpha$, since the drift term of the FPE must push all probability from  the neighborhood towards this maximum, as we know from the deterministic limit. Conversely, minima of $p_s$ are associated with completely unstable fixed points of $\vec \alpha$. The remaining stationary points of $p_s$ are saddles. Since $\vec j_s$ must also follow the height lines of $p_s$ near saddles, it follows that the fixed point of $\vec j_s$ at the saddle of $p_s$ is a saddle point, which has one attractive and one repulsive eigendirection. Concordantly, $\vec \alpha$ also has a saddle point at this position in state space. The reason is that the direction in which $p_s$ decreases/increases must be a stable/unstable direction of $\vec f$ so that probability is pushed away from that fixed point in the direction in which $p_s$ increases, and toward that fixed point along the direction in which $p_s$ decreases.  

To give an explicit example, we consider in the following the case that $ \mathbf{D} \propto  \text{Id}$, and we linearize the convective field around its fixed point, which we place without loss of generality at the origin, 
\begin{equation}
    \vec{\alpha} = \begin{pmatrix}
    ax + by\\
    cx + ey
    \end{pmatrix}.
\end{equation}

The Jacobian reads
\begin{equation}
    J_{\vec{\alpha}} = \begin{pmatrix}
    a & b\\
    c & e    
\end{pmatrix}
\label{eq:alpha:lin_ansatz}
\end{equation}
and has the eigenvalues
\begin{equation}\label{lambda+-}
    \lambda_{\pm} = \frac{1}{2}\left(-\tau \pm \sqrt{\tau^2 - 4 \Delta}\right)
\end{equation}
with $\tau = \text{tr}(J_{\vec{\alpha}}) = a + e$ and $\Delta = \text{det}(J_{\vec{\alpha}}) = ae-bc$.

Furthermore we can calculate $\vec{\alpha}_{\parallel}$ and $\vec{\alpha}_{\perp}$, which yields
\begin{equation}
    \vec{\alpha}_{\parallel} = \begin{pmatrix}
    \frac{a + e}{\eta}\left(\Pi y + \left(\Delta + a^2 + c^2 \right)x\right)\\
    \frac{a + e}{\eta}\left(\Pi x + \left(\Delta + b^2 + e^2\right)y\right)
    \end{pmatrix}
\end{equation}
\begin{equation}
    \vec{\alpha}_{\perp} = \begin{pmatrix}
    \phantom{+}\frac{b - c}{\eta}\left(\Pi x + \left(\Delta + b^2+ e^2\right)y\right)\\
    - \frac{b - c}{\eta}\left(\Pi y + \left(\Delta + a^2 + c^2 \right)x\right)
    \end{pmatrix}
\end{equation}
with $\eta = (b - c)^2 + (a + e)^2$, $\Pi = ab + ce$ and $\Delta = \text{det}(J_{\alpha}) = ae - bc$.

Since in leading order $\vec{j}_s = \vec{\alpha}_{\perp} \cdot p_s$, 
we can write the Jacobian of $\vec{j}_s$ as
$J(\vec{j}_s) = J(\vec{\alpha}_{\perp})\cdot p + R$
where every entry of the rest matrix $R$ is proportional to a derivative of $p_s$ and thus vanishes at the location of the fixed point.
This means that the type of a fixed point of $\vec{j}_s$ is always identical to the type of the corresponding fixed point of $\vec{\alpha}_{\perp}$.

The Jacobian of $\vec{\alpha}_{\perp}$ is given by
\begin{equation}
    J_{\vec{\alpha}_{\perp}} = \frac{1}{\eta}\begin{pmatrix}
    (b-c)\Pi \phantom{empty} (b-c)\left(\Delta + b^2 + e^2\right)\\
    - (b-c) \left(\Delta + a^2 + c^2\right) \phantom{em} - (b-c) \Pi
    \end{pmatrix}
\end{equation}
and has the eigenvalues
\begin{equation}
    \mu_{\pm} = \pm \frac{(b-c) \Delta}{\sqrt{-\Delta \eta}}\, .
\end{equation}

Since $\eta > 0$, these eigenvalues are either two real values with opposite signs, or a complex conjugate pair with vanishing real part. The first case occurs for $\Delta < 0$ and implies a saddle point in $\vec{\alpha}_{\perp}$ and therefore in $\vec{j}_s$, and from eq.~\eqref{lambda+-} we know that this  corresponds to a saddle point in $\vec{\alpha}$. The second case occurs for $\Delta > 0$ and implies a center in $\vec{\alpha}_{\perp}$ and thus in $\vec{j}_s$. All fixed points of $\vec{\alpha}$ that are not saddle points are associated with centers of $\vec j_s$.

\smallskip

For the special case $b = c$, which can also be expressed as the more general condition $\vec{\nabla}\times\mathbf{D}^{-1}\vec{\alpha}(\vec{x}_0) = 0$, both eigenvalues are zero. We will see below that in this case the leading order contributions in $1/N$ vanish and the fixed points of $\vec{\alpha}$ and $\vec{j}_s$ no longer coincide. Instead, the stationary currents become dipole-like.

\subsection{The rotation sense of $\vec{j}_s$}
The curl of any vector field $\vec f(\vec x)$ at location $\vec{x_0}$ can always be split into two distinct contributions: One contribution stems from the circular flow of the vector field around $\vec x_0$, the other one from the change of its absolute value in the vicinity of $\vec{x_0}$. For this purpose we write $\vec f(\vec{x}) = |f| \cdot \vec{e}_f$ and calculate
\begin{equation}
\vec{\nabla}\times \vec{f} = \vec{\nabla}\times\vec{e}_f \cdot |f| + \vec{\nabla} |f| \times \vec{e}_f
\end{equation}
At the location of a fixed point of $\vec{f}(\vec{x})$ the second term vanishes and the curl is given alone by the circular flow around the fixed point. We can then interpret the sign of this first term as the rotation sense of $\vec{f}$ around that fixed point.

Thus, in order to obtain information about the rotation sense of $\vec j_s$, we first derive properties of its curl. 
From equation \eqref{eq:fpe:j_definition} we obtain
\begin{align}
\begin{aligned}
    \vec{\nabla}\times \mathbf{D}^{-1}\vec{j}_s &= 
   \vec{\nabla}\times \left(\mathbf{D}^{-1} \vec{\alpha} p_s - \frac{1}{2N} \vec{\nabla}p_s\right)\\
   &= \vec{\nabla}\times \left(\mathbf{D}^{-1} \vec{\alpha}  p_s\right)\\
    \end{aligned}
    \label{eq:rotj_general}
\end{align}

At an extremum of $p_s$, the gradient of $p_s$ vanishes, and this relation simplifies to 
 \begin{eqnarray}
   \text{sgn} \left( \vec{\nabla}\times \mathbf{D}^{-1}\vec{j}_s\right)
 &=& \text{sgn} \left( \vec{\nabla}\times \mathbf{D}^{-1}\vec{\alpha}\right) \label{eq:rot_sinn_j_allg}\\
  &=& \text{sgn} \left( \vec{\nabla}\times \mathbf{D}^{-1}\vec{\alpha}_\perp\right) .\label{eq:rot_sinn_j}
\end{eqnarray}
In the last step, we have used equations \eqref{eq:alphazerlegung} and \eqref{eq:def:alpha_parallel}.
There is no simple way to obtain from here $\text{sgn} \left( \vec{\nabla}\times \vec{j}_s\right)$, which gives at an extremum of $p_s$ the rotation sense of $\vec j_s$. In many cases, however, we expect that $\text{sgn} \left( \vec{\nabla}\times \mathbf{D}^{-1}\vec{j}_s\right) = \text{sgn} \left( \vec{\nabla}\times \vec{j}_s\right)$.

In the following, we discuss two cases in which we can obtain information directly about $\text{sgn} \left( \vec{\nabla}\times \vec{j}_s\right)$:
The first case is that of large $N$, where we obtain from relation \eqref{eq:j:leading_order} at an extremum of $p_s$
\begin{equation}
     \text{sgn} \left( \vec{\nabla}\times \vec{j}_s\right)
  = \text{sgn} \left( \vec{\nabla}\times \vec{\alpha}_\perp\right) \label{eq:rot_sinn_j_ohne_D} \, .
\end{equation}
In this case, we need to  find the component $\vec\alpha_\perp$ of the convective field and to evaluate the sign of its curl. 

The second case is that of constant, isotropic diffusion, 
\begin{equation}
    \mathbf{D} = d \cdot \text{Id}\, . \label{eq:D=Id}
\end{equation}
In this case, we obtain from equations \eqref{eq:rot_sinn_j_allg} and \eqref{eq:alpha}
\begin{equation}
     \text{sgn} \left( \vec{\nabla}\times \vec{j}_s\right)
  = \text{sgn} \left( \vec{\nabla}\times \vec{\alpha}\right) =  \text{sgn} \left( \vec{\nabla}\times \vec{f}\right)\, ,
  \label{eq:rot_sinn_j_d_const_leading_N}
\end{equation}
and we need not determine the two components of $\vec\alpha$.

Away from an extremum of $p_s$, the sign of the curl is not necessarily identical to the rotation sense of $\vec j_s$. This can be seen most easily in the  case \eqref{eq:D=Id} of constant, isotropic diffusion, for which eq.~\eqref{eq:rotj_general} simplifies to
\begin{equation}
    \vec{\nabla}\times\vec{j}_s= \vec{\nabla}\times \vec{f} \cdot p_s + \vec{\nabla}p_s \times \vec{f}.\\
    \label{eq:rotj_isotropic}
\end{equation}
The sign of the first term gives the rotation sense of $j_s$ around the extremum of $p_s$. 

The  second term of equation \eqref{eq:rotj_isotropic} does not vanish outside extrema of $p_s$.  For negative $\vec\nabla p_s$, i.e., in the vicinity of maxima of $p_s$,  the second term points in the opposite direction of the first term. Close to minima, it points in the same direction. This second term can cause a difference in the sign of the curl of $j_s$ and the sense in which $j_s$ circles around the extremum of $p_s$. 

In order to gain a deeper understanding of all this, let us consider the example of a rotationally symmetric model with constant, isotropic diffusion \eqref{eq:D=Id}, which we will solve explicitly  in the limit of large $N$.

\subsubsection*{Example: Rotationally symmetric model}

If the vector field $\vec f(\vec{x})$ obeys a rotational symmetry around its fixed point, it has the form
\begin{equation}\label{eq:rotsymmmodel1}
\vec f(\vec{x}) = f_{r}(r) \vec{e}_{r} + f_{\varphi}(r) \vec{e}_{\varphi}\, .
\end{equation}
Choosing \eqref{eq:D=Id} for the diffusion matrix, we have $\vec f = \vec \alpha$ and
\begin{equation}\label{eq:rotsymmmodel2}
 f_{r}(r) \vec{e}_{r}= d \cdot \vec \alpha_\parallel \quad \text{ and } \quad f_{\varphi}(r) \vec{e}_{\varphi} =\vec \alpha_\perp  \, .
\end{equation}
From relation \eqref{eq:analytisch:p_solution}, which is valid for large $N$,  we obtain the stationary probability distribution
\begin{equation}
    p_s = e^{\frac{2N}{d}\int f_r(r)\text d r}\, . \label{eq:ps_rotsymm}
\end{equation}

Furthermore equation \eqref{eq:j:leading_order} gives 
\begin{equation}
    \vec{j}_s = f_{\varphi}(r) p_s \vec{e}_{\varphi} , \label{eq:js_rotsymm}
\end{equation}
i.e. $\vec{j}_s$ goes in circles around the fixed point of $\vec{f}$, and its radial component vanishes.
The curl of $\vec{j}_s$ reads then
\begin{align}
    \left(\vec{\nabla}\times\vec{j}_s\right)_z &= \underbrace{\left(f_{\varphi}'(r) + \frac{f_{\varphi}(r)}{r}\right)}_{\vec{\nabla}\times\vec{f}} p_s(r) + \underbrace{f_{\varphi} (r) p'_s(r)}_{\vec{\nabla}p_s \times \vec{f}}
    \label{eq:rotj:rotationssym:form1}
    \\
    &= \underbrace{\frac{1}{r} \cdot f_{\varphi}(r) p_s(r)}_{\vec{\nabla}\times\vec{e}_j \cdot |j_s|}  + \underbrace{\left(f_{\varphi}' (r) p_s(r) + f_{\varphi} (r) p'_s(r) \right)}_{\vec{\nabla}|j_s| \times \vec{e}_j}
    \label{eq:rotj:rotationssym:form2}
\end{align}
with the first term describing a circular flow around the fixed point and the second term being a a less intuitive contribution induced by the change of the absolute value of $\vec{j}_s$.
At the maximum of $p_s$, we have only the first term, and the rotation sense of $\vec j_s$ is given directly by the sign of $f_\varphi$.

When $N$ is large, most of the stationary probability density is located close to the center, and one can simplify these expressions further by approximating $\vec f$ by linear terms in $x$ and $y$, 
\begin{equation}
    \vec{f} = \begin{pmatrix}
    -\lambda x - \omega y\\
    -\lambda y + \omega x
    \end{pmatrix}.
    \label{eq:example:spirale}
\end{equation}
This means that we have set $f_r=-\lambda r$ and $f_\varphi = \omega r$. 

The  eigenvalues of the Jacobian at the fixed point
read $\{-\lambda - i \omega, - \lambda + i \omega\}$, i.e. the fixed point is a stable spiral for $\lambda > 0, \omega \neq 0$, with the sign of $\omega$ determining the sense of the spiral. For $\omega = 0$ the fixed point is a stable node.
\smallskip

\begin{figure}[H]
\subfigure[]{\includegraphics[width=0.48\linewidth]{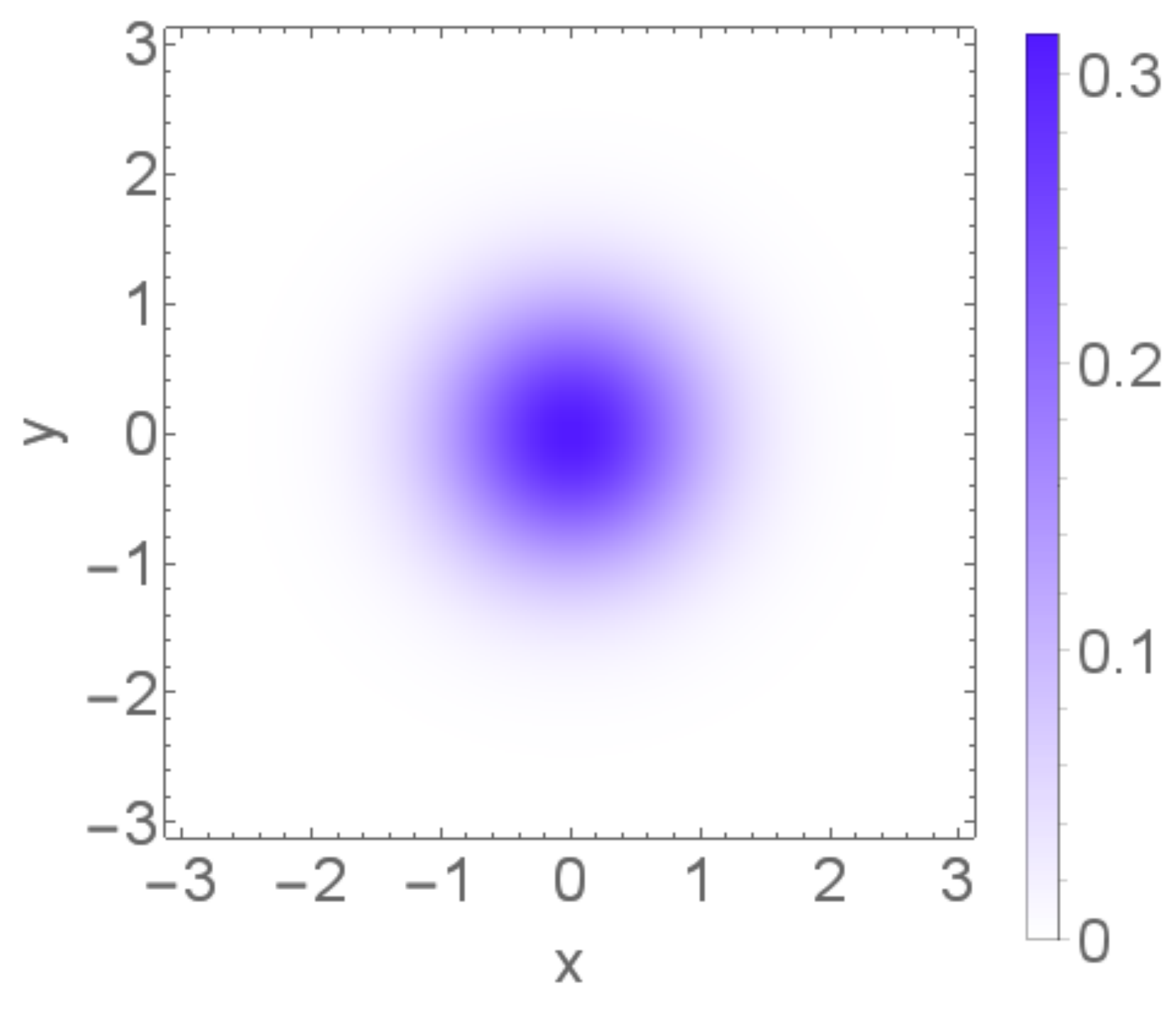}}
\subfigure[]{\includegraphics[width=0.4\linewidth]{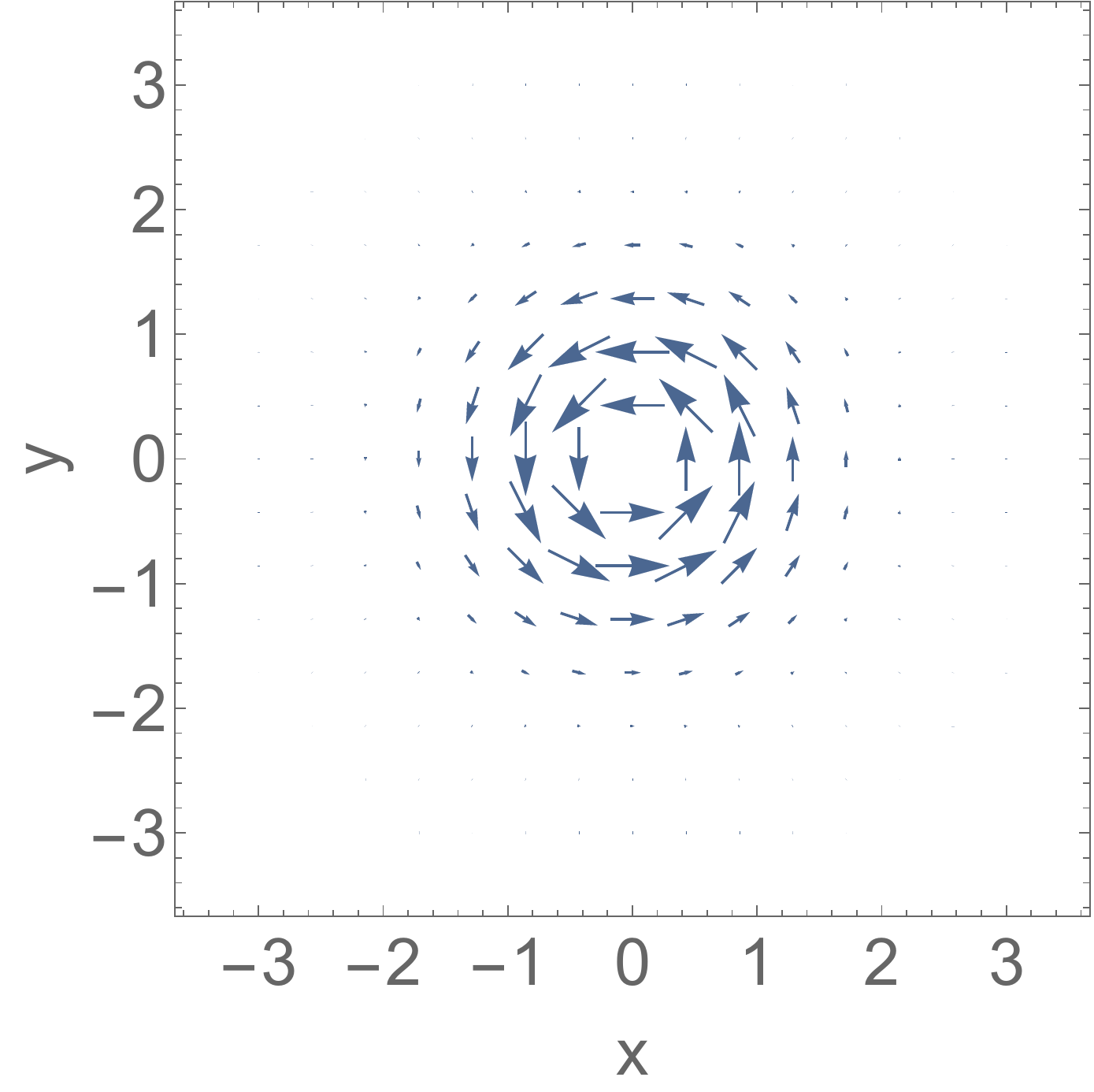}}\\
\subfigure[]{\includegraphics[width=0.48\linewidth]{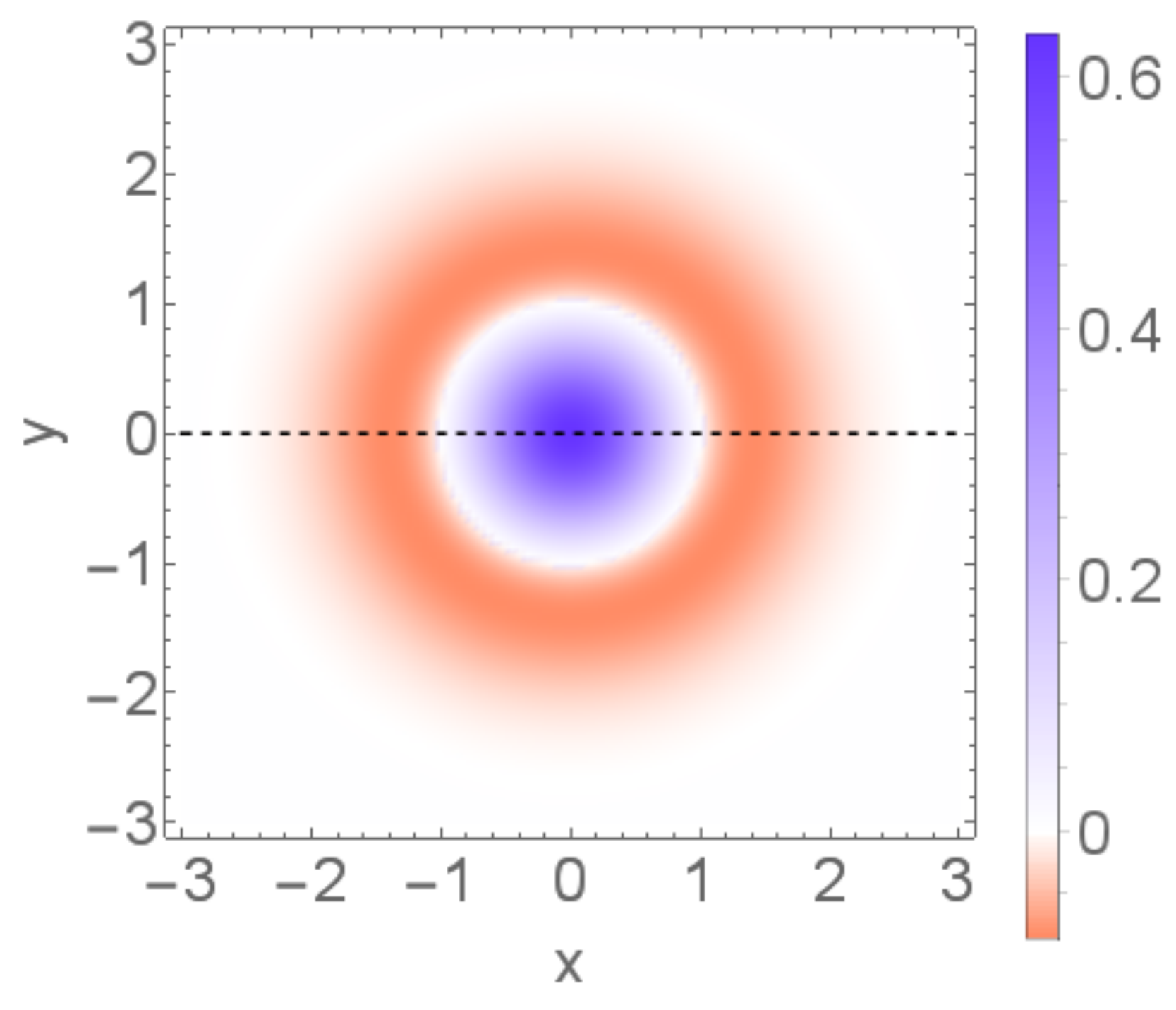}}
\subfigure[]{\includegraphics[width=0.51\linewidth]{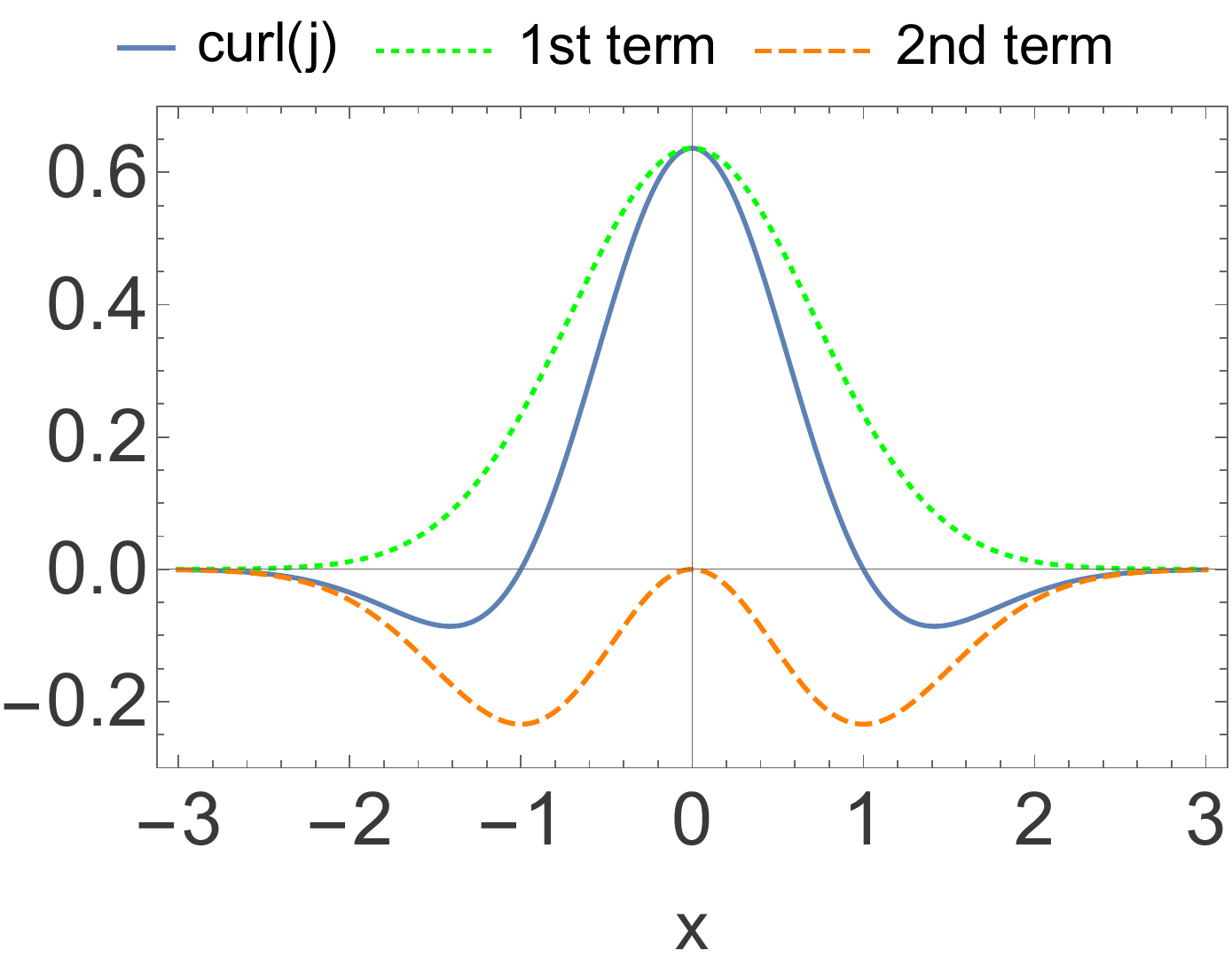}}
\caption{(Color online) The different contributions to $\vec j_s$ for the rotationally symmetric system \eqref{eq:example:spirale} for the parameter values $\omega = \lambda = d = N= 1$:\\
(a) stationary probability density $p_s(x,y)$\\
(b) stationary probability current $\vec{j}_s (x,y)$\\
(c) curl of the stationary probability current $|\vec{\nabla} \times \vec{j}_s(x,y)|$ (as all vectors point along the positive $z$-axis, only the absolute value is shown)\\
(d) cut along the x-axes in (c)
\label{fig:j_contributions}}
\end{figure}

We obtain furthermore
\begin{equation}
\vec{\nabla}\times \vec{f} = 2 \omega
\end{equation}
and, using \eqref{eq:ps_rotsymm}, 
\begin{equation}
    p_s(x) = A \exp\left(-\frac{\lambda N}{d}(x^2+y^2)\right).
\end{equation}
The stationary current \eqref{eq:js_rotsymm} is then
\begin{equation}
    \vec{j}_s(x) = p_s \begin{pmatrix}
     -\omega y\\
     \omega x
    \end{pmatrix},
\end{equation}
giving
\begin{align}
\begin{aligned}
    \left(\vec{\nabla}\times \vec{j}_s(x)\right)_z &= \underbrace{2 \omega p_s}_{\text{1st term}} + \underbrace{- \frac{\omega \lambda N}{d} p_s (x^2+y^2)}_{\text{2nd term}}\\
    &= \frac{\omega\lambda N}{d} p_s  \left(\frac{2 d}{\lambda N} - (x^2+y^2)\right).
\end{aligned}
\end{align}
If we choose $\omega > 0$, the curl of $\vec j_s$  is positive inside a circle with radius $r = \sqrt{\frac{2d}{\lambda N}}$ around the origin where the first term, which gives the rotation sense of $\vec j_s$ around its fixed point, dominates and negative for larger distances from the origin. These different quantities are illustrated  in fig.~\ref{fig:j_contributions}.

\subsection{The general topology of $\vec{j}_s$}
\label{sec:predict:j}

Taking the previous results together, we have established a method for finding the fixed points of $\vec{j}_s$ in leading order in $1/N$ and to calculate the sense in which $j_s$  flows around maxima of the stationary probability distribution from \eqref{eq:rot_sinn_j_ohne_D}, or, in the case of constant, isotropic diffusion, from the simpler relation \eqref{eq:rot_sinn_j_d_const_leading_N}.

In the following, we explore the usefulness and limits of the large-$N$ approximation for artificial models, where only some of the simplest possible coupling terms between the two chemical species are included.

An uncoupled two-dimensional reaction system can be written in the form
\begin{equation}
    \vec{f^u} = \begin{pmatrix}
    f^u_1(x)\\
    f^u_2(y)
    \end{pmatrix}
    \label{eq:uncoupled:drift}
\end{equation}
and
\begin{align}
\mathbf{D} = \begin{pmatrix}
D_{11}(x) & 0\\
0 & D_{22}(y)
\end{pmatrix}.
\label{eq:uncoupled:diffusion}
\end{align}

If we now were to introduce coupling reactions into this system, we would obtain coupling terms inside both the drift vector and the diffusion matrix. In order to understand the effect of each of these terms separately, we will first consider artificial models where only the drift vector or the diffusion matrix contains coupling terms. In a following step we will then consider realistic models. 

\subsubsection{$\vec{j}_s$ for coupling terms inside the deterministic drift}

There are many different ways to introduce coupling terms into the deterministic drift of \eqref{eq:uncoupled:drift}. However not all types of coupling terms can be produced by chemical reactions. For example, a positive linear coupling in the form $f_1(x, y) = f^u_1(x) + a y$ can occur due to a reaction $Y \overset{a}{\longrightarrow} X + Y$. 
But a negative linear coupling like $f_1(x, y) = f^u_1(x) - a y$ is not possible, as it would lead to negative numbers of $X$ molecules.  

\smallskip
Considering restrictions like this, the most general form of the deterministic drift of a coupled two-species reaction system with up to bimolecular interactions and mass-action kinetics reads
\begin{equation}
    \vec{f} = \begin{pmatrix}
    f^u_1(x) + a_1 y + b_1 x y + c_1 y^2\\
    f^u_2(y) + a_2 x + b_2 x y + c_2 x^2\\
    \end{pmatrix}
    \label{eq:drift:modified}
\end{equation}
with $a_i, c_i > 0$ and $b_i \in \mathbb{R}$.
The coupling constants can be independent from each other as it is possible to construct a chemical reaction for any of these coupling terms that does not alter the deterministic drift in any other way.

\smallskip
Using \eqref{eq:uncoupled:diffusion} and \eqref{eq:drift:modified} we obtain the following explicit expression for \eqref{eq:rot_sinn_j_allg}, which is valid at an extremum of $p_s$:
\begin{align}
\left(\vec{\nabla}\times \mathbf{D}^{-1} \vec{j}_s \right)_z 
= \left(\vec{\nabla}\times \mathbf{D}^{-1} \vec{f} \right)_z \, ,
\label{eq:drift:prediction:common}
\end{align}
or, equivalently,
\begin{align}
\frac{\partial_x j_y}{D_{22}} - \frac{\partial_yj_x}{D_{11}}
= \frac{\partial_x f_y}{D_{22}} - \frac{\partial_yf_x}{D_{11}}\, .
\label{eq:drift:prediction:common2}
\end{align}

Unfortunately, the sign of $\vec{\nabla}\times \vec{j}_s $ at the considered extremum of $p_s$ cannot be deduced with certainty from this expression.
To leading order in $1/N$ we have the alternative route via calculating  $\text{sign}(\vec{\nabla}\times\mathbf{D}^{-1}\vec{\alpha}_{\perp})$, but it is impossible to write down a general analytical expression for $\vec{\alpha}_{\perp}$ for this system. 

We therefore confine ourselves again to the case 
\begin{align}
\mathbf{D} = \begin{pmatrix}
d & 0\\
0 & d
\end{pmatrix},
\label{eq:fp_example:diffusion}
\end{align}
for which $D_{11}=D_{22}=d$.

Setting \eqref{eq:drift:prediction:common2} to zero yields an implicit expression for the curve that divides the two regions with opposite signs for the curl of $\vec f$, which equals the curl of $\vec j$. We  take it in the following as an estimate for the rotation sense of $\vec j$ around its fixed point. Since both $x$ and $y$ occur only linearly in this equation, this curve must be a line. For $c_1 \neq b_2$ we can write it in the explicit form 

\begin{equation}
    y = \frac{2 {c}_2 - {b}_1}{2 {c}_1 - {b}_2} x + \frac{{a}_2 - {a}_1}{2 {c}_1 - {b}_2}\, .
    \label{eq:boundary_line}
\end{equation}
Depending on the parameters of the system the distance of this boundary line from the maximum of $p_s$ will be smaller or larger.

\begin{figure}[h]
\hfill
\subfigure[]{\includegraphics[width=0.4\linewidth]{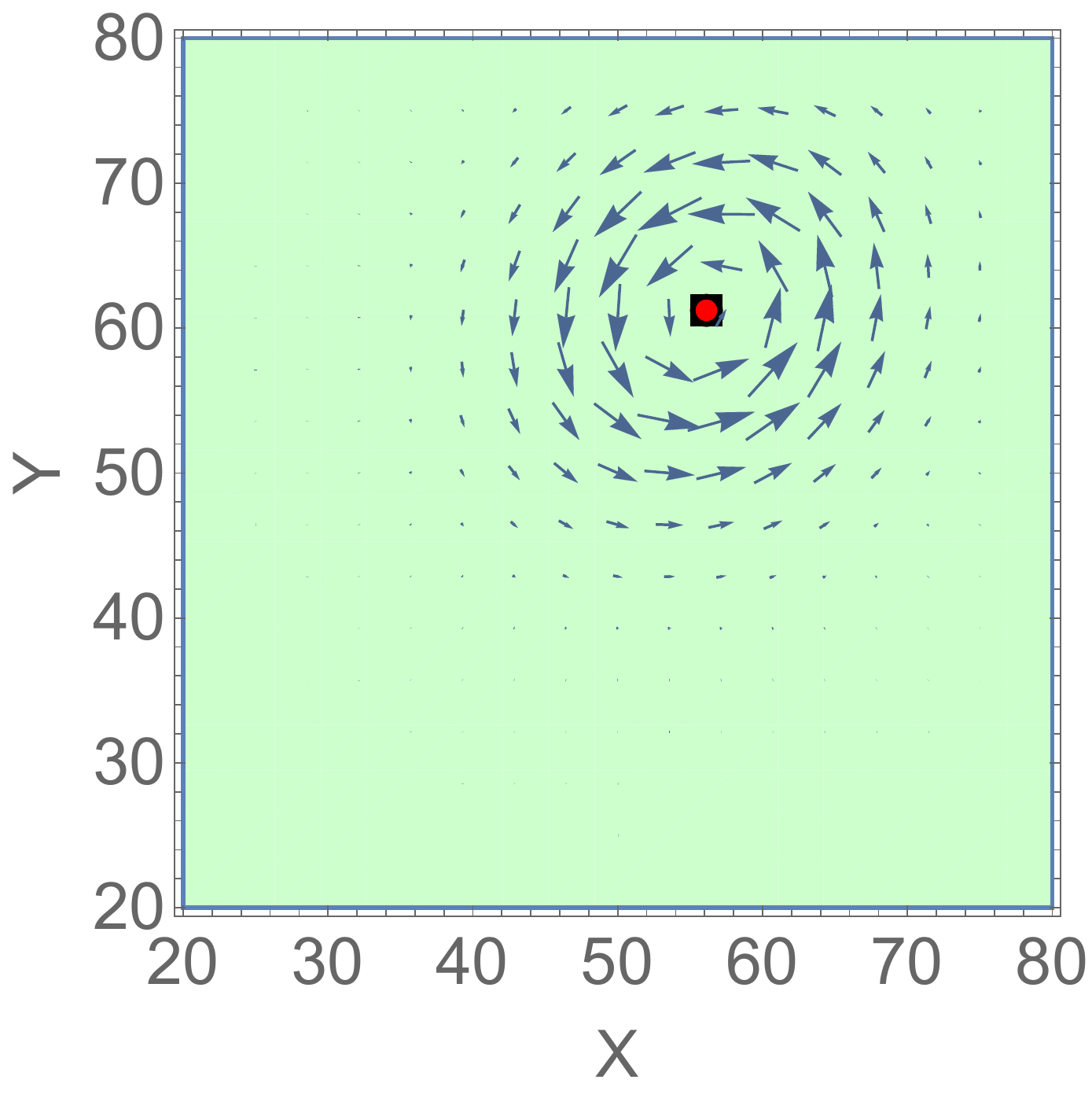}}
\hfill
\subfigure[]{\includegraphics[width=0.4\linewidth]{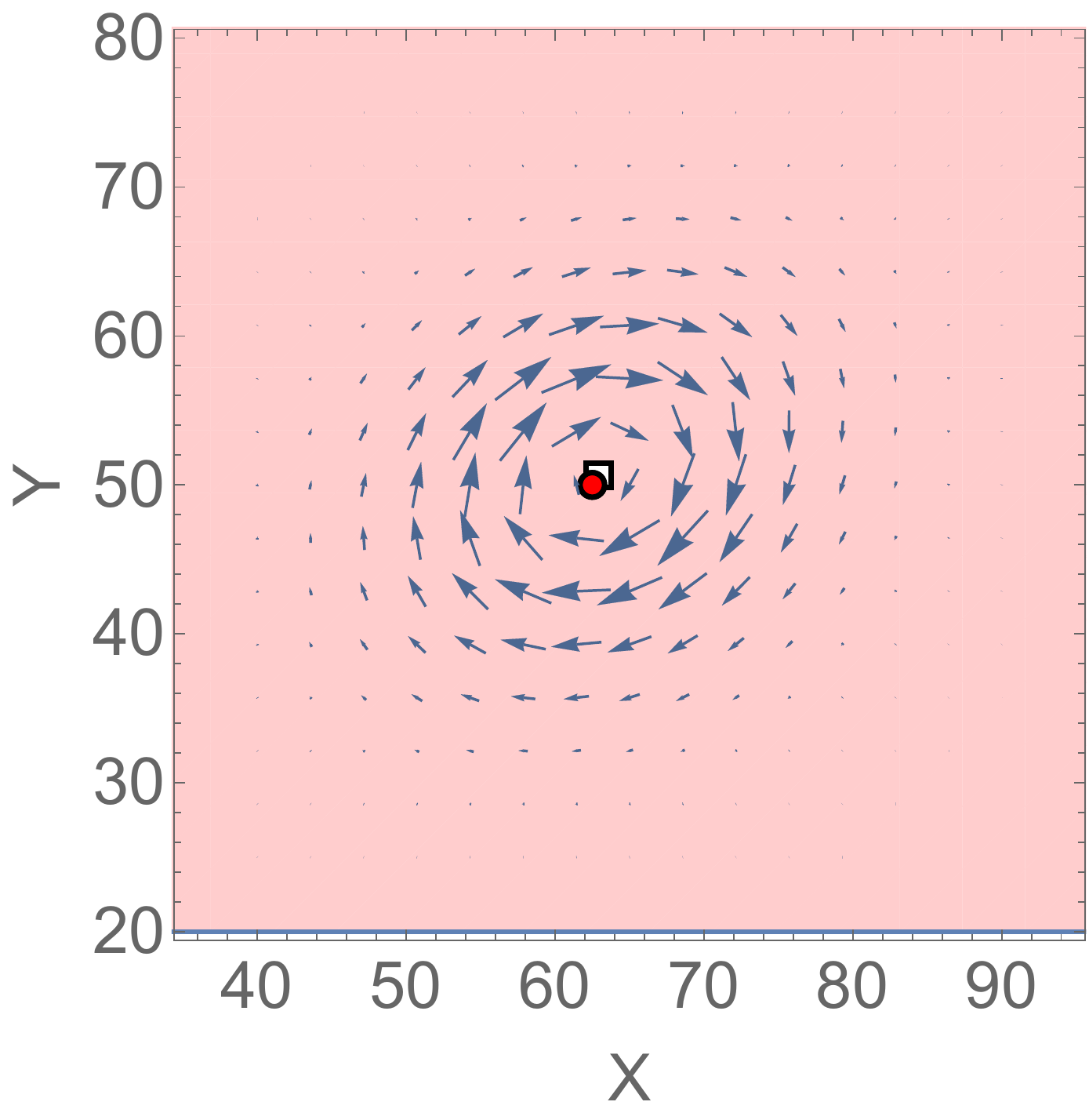}}
\hfill

\hfill
\subfigure[]{\includegraphics[width=0.4\linewidth]{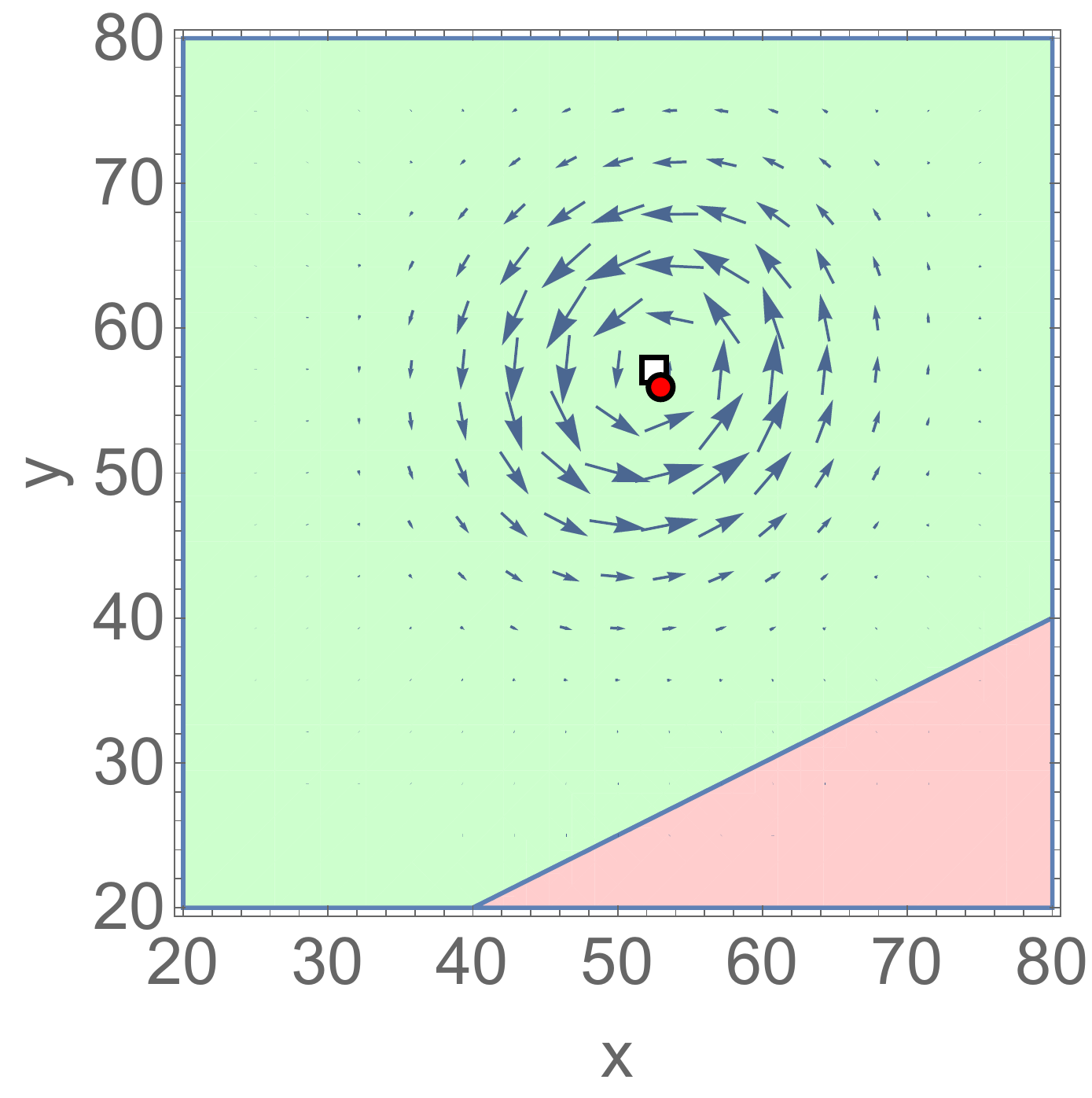}}
\hfill
\subfigure[]{\includegraphics[width=0.4\linewidth]{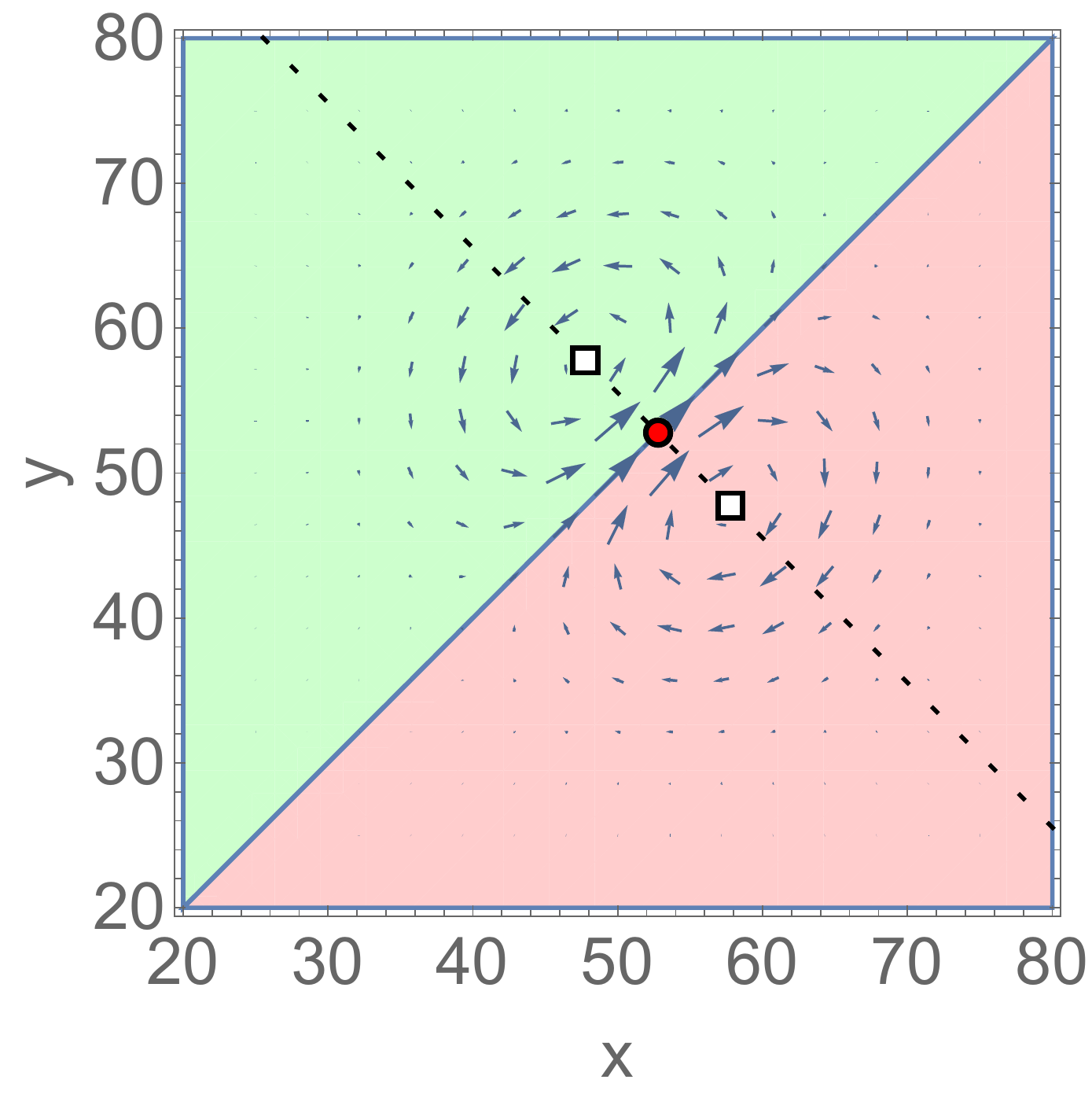}}
\hfill

\caption{
\label{fig:coupl:drift}
(Color online) Stationary probability current $\vec{j}_s$ for the coupled system \eqref{eq:drift:modified} with \eqref{eq:fp_example:drift}, \eqref{eq:fp_example:diffusion} for the parameter set $r_1 = r_2 = 50$, $s_1 = s_2 = 1, d = 100, N = 1$, and different values for the coupling terms: \\
(a) $a_1 = 0.1, a_2 = 0.2, b_i = c_i = 0$\\
(b) $a_i = b_i = c_2 = 0, c_1 = 0.005$\\
(c) $a_i = c_i = 0, b_1 = 0.001; b_2=0.002$\\
(d) $a_i = c_i = 0, b_1 = b_2 = 0.001$\\
The green (lighter)/red (darker) background color indicates the predicted positive/negative rotation sense as of eq.~\eqref{eq:drift:prediction:common}, i.e., based on the sign of the curl of $\vec \alpha$. The red dots indicate the position of the fixed point of $\vec{\alpha}$, the white rectangles the fixed points of $\vec{j}_s$.
}
\end{figure}

Fig.~\ref{fig:coupl:drift} shows some example systems where the estimated rotation sense of $\vec{j}_s$ according to \eqref{eq:drift:prediction:common} is plotted as a green (positive) resp. red (negative) background color together with the fixed points of $\vec{\alpha}$ (red) and overlayed with vector plots of $\vec{j}_s$ from numerical solutions of the Fokker-Planck equation. 

For this figure, we used 
\begin{align}
\vec{f}^u = \begin{pmatrix}
r_1 - s_1 x\\
r_2 - s_2 y
\end{pmatrix}
\label{eq:fp_example:drift}
\end{align}
as the underlying uncoupled system and the diffusion marix \eqref{eq:fp_example:diffusion}.

\medskip

Fig.~\ref{fig:coupl:drift} shows that
if the fixed point of $\vec{\alpha}$ is sufficiently far away from the boundary line, the fixed points of $\vec{j}_s$ match the $\vec{\alpha}$ fixed points very well. For these systems the topology of $\vec{j}_s$ can  be estimated based an these fixed points and the curl of $\vec \alpha$ alone. Only for more detailed properties of $\vec{j}_s$, like the length of the $\vec{j}-$vectors or the elliptical deformation of the circular currents around the fixed point, is it necessary to solve the Fokker-Planck equation and calculate $\vec{j}_s$ explicitly.

\smallskip
The figure also shows one shortcoming of our method: While we can predict the rotation sense and in leading order in $1/N$ also the fixed points of $\vec{j}_s$, we have no information about the strength of the stationary probability currents. For example, we could not foresee how tiny the values of $\vec{j}_s$ in the region with negative rotation sense in fig.~\ref{fig:coupl:drift} (c) are.
\medskip

The situation is qualitatively different in  fig.~\ref{fig:coupl:drift} (d), for which $\vec{\nabla}\times\mathbf{D}^{-1}\vec{\alpha} = 0$ at the fixed point of $\vec\alpha$. The shape of $\vec j_s$ can therefore no longer be obtained by analysing the sign of $\vec{\nabla}\times\mathbf{D}^{-1}\vec{\alpha}$. 
As fig.~\ref{fig:coupl:drift} (d) shows, $\vec{j}_s$ assumes a dipole-like shape in this scenario. 

While the discrepancy between the $\vec{\alpha}$ and $\vec{j}_s$ fixed points in fig.~\ref{fig:coupl:drift}(d) is obvious, the maximum of $p_s$ and the fixed point of $\vec{\alpha}$ have also become different, even though their distance is very small and not visible in the figure. Both these points lie exactly on the $\vec{\nabla}\times\mathbf{D}^{-1}\vec{\alpha} = 0$ boundary line due to the symmetry of the system.

In the following, we will analyse the emergence of dipole currents in more detail.

\smallskip
\subsubsection{How higher-order terms create dipole currents}
 
In \eqref{eq:j:leading_order} we have seen that the stationary probability current in leading order in $1/N$ is given by
\begin{equation}
\vec{j}_s = \vec{\alpha}_{\perp} p_s \nonumber \, .
\end{equation}

In this approximation $\vec{\alpha}_{\perp}$ has no influence on $p_s$, and fixed points of $\vec{\alpha}$ and $\vec{j}_s$ and extrema of $p_s$ coincide.

However if this leading order term vanishes in the neighborhood $V(\vec{x_0})$ of a point $\vec{x}_0$, higher order terms in $1/N$ determine the shape of $\vec j_s$. We can than write for $\vec{x} \in V(\vec{x_0})$
\begin{align}
\begin{aligned}
& \vec{\alpha}_{\perp} (\vec{x}) = 0\\
\Rightarrow & \vec{\alpha} (\vec{x}) = \mathbf{D}\vec{\alpha}_{\parallel} (\vec{x})\\
\Rightarrow & \vec{\nabla}\times\mathbf{D}^{-1}\vec{\alpha} ( \vec{x}) = 0
\end{aligned}
\label{eq:higher_order_relevant}
\end{align}
which means that $\vec{x}$ lies on the boundary line between positive and negative rotation sense. Consequently, the significance of higher order terms in $1/N$ near a fixed point  of $\vec{\alpha}$ is given by the proximity of that fixed point to the boundary line.

\medskip

The observation that the emerging higher order currents are dipole shaped is plausible as a dipole current is the simplest current with vanishing curl along a line. We can understand this result better by using an example. Let us assume a single stable fixed point of $\vec f$  located at the origin. Since we are interested in the current in the vicinity of this fixed point, we choose a model linear in $x$ and $y$, 
\begin{equation}
    \vec{f} = \begin{pmatrix}
    -x\\
    -y
    \end{pmatrix}.
\end{equation}
For the diffusion matrix we choose one of the simplest possible forms that gives a nonzero stationary current,
\begin{equation}
    \mathbf{D} = \begin{pmatrix}
    d & 0\\
    0 & d + x
    \end{pmatrix}.
\end{equation}
For this simple model, we have  $\vec{\alpha} = \vec{f}$ and
\begin{equation}
    \vec{\nabla}\times\mathbf{D}^{-1}\vec{\alpha} = -\frac{y}{(d+x)^2} \vec{e}_z\, ,
\end{equation}
which is indeed zero at the location of the fixed point.

We find the stationary probability density $p_s$ close to the fixed point, i.e., for small $x$ and $y$ by an educated guess,
\begin{equation}
    p_s = A (d-x) \exp\left(-\frac{N}{d}(x^2 + y^2)\right)
    \label{eq:dipol:analytisch:ansatz}
\end{equation}
Indeed, this proves to be a stationary solution to leading order of $x$ and $y$: 
\begin{align}
    \begin{aligned}
        \vec{\nabla} \cdot \vec{j} =& \frac{\partial}{\partial x} \left(-x p_s - \frac{d}{2N} \frac{\partial p_s}{\partial x}\right)\\
        +& \frac{\partial}{\partial y} \left(-y p_s - \frac{d-x}{2N} \frac{\partial p_s}{\partial y}\right)\\
         =& A \left(-\frac{d}{2N}\frac{\partial}{\partial x}  - x \right)e^{-\frac{N}{d}(x^2 + y^2)} + \mathcal{O}(x^2, y^2)\\
        =& 0 + \mathcal{O}(x^2, y^2)\, .
    \end{aligned}
\end{align}
Thus the stationary probability current close to the $\vec{f}$ fixed point is approximated by
\begin{equation}
    \vec{j_s} = \begin{pmatrix}
    - xp - \frac{d}{2N} \frac{\partial p}{\partial x}\\
    - yp - \frac{d - x}{2N} \frac{\partial p}{\partial y}
    \end{pmatrix} \approx
    \begin{pmatrix}
    \frac{d A}{2N}\\
    0
    \end{pmatrix} + \mathcal{O}(x, y)\, .
\end{equation}
This constant vector pointing along the $x$-axis is the first-order term of a dipole field that points in the $x$  direction.

\smallskip

The fact that such a dipole field arises can also be seen from the Ansatz \eqref{eq:dipol:analytisch:ansatz}, which is the leading order result of the summation of two equal Gaussian distributions which are $1/d$ apart from each other.

\smallskip

\subsubsection{$\vec{j}_s$ for coupling terms inside the diffusion matrix}

Let us now investigate models with coupling terms inside the diffusion matrix. Again, in order to understand the influence of these coupling terms in greater detail, we will not consider realistic coupling reactions but assume a completely uncoupled drift vector $\vec f^u$, as given in \eqref{eq:uncoupled:drift}.

The general uncoupled diffusion matrix \eqref{eq:uncoupled:diffusion} permits two ways of introducing a coupling: Either the elements $D_{ii}(x_i)$ become dependent on the respective other species $X_j$, or the off-diagonal terms $D_{12} = D_{21}$ become non-zero.

Just as the coupling terms inside the deterministic drift, couplings inside the diffusion matrix must obey certain restrictions when they should be producible by mass-action kinetics. Especially, since the $D_{ii}(x_i)$ can never decrease due to an additional interaction, all introduced coupling terms on the diagonal of $\mathbf{D}$ must be positive. 

The off-diagonal terms on the other hand may have either positive or negative sign, except for the constant term which is always positive.

\smallskip

All in all we arrive at the following general coupled diffusion matrix elements:
\begin{align}
\begin{aligned}
D_{11}(x,y) =& D_{11}^u(x) + a_1 y + b_1 x y + c_1 y^2\\
D_{22}(x,y) =& D_{22}^u(y) + a_2 x + b_2 x y + c_2 x^2\\
D_{12}(x,y) =& D_{21}c(x,y) = e \pm f_1 x \pm f_2 y\\
&\pm g x y \pm h_1 x^2 \pm h_2 y^2
\end{aligned}
\end{align}
for reaction systems with mass-action kinetics and up to bimolecular reactions.

\medskip

We focus now on a simplified example with only linear diagonal coupling terms given by
\begin{equation}
    \mathbf{D} = \begin{pmatrix}
    D_{11}^u (x) + a_1 y & 0 \\
    0 & D_{22}^u (y) + a_2 x
    \end{pmatrix}.
    \label{eq:diff:diag:linear}
\end{equation}

With this expression for the diffusion matrix, we obtain
\begin{align}
\begin{aligned}
&\vec{\nabla}\times\mathbf{D}^{-1}\vec{\alpha} = 
     \frac{a_1(- 2 f_1(x) - D^{'u}_{11}(x))}{2(a_1 y + D^u_{11}(x))^2}\\
     &- \frac{a_2(- 2 f_2(y) - D^{'u}_{22}(y))}{2(a_2 x + D^u_{22}(y))^2}\\
     &=\phantom{-} \alpha_1(x)~ a_1 ~ \eta_1^2(x, y) - \alpha_2(y)~ a_2 ~ \eta_2^2(x,y)
\end{aligned}
\label{eq:diff:lin:rotj}
\end{align}
with the abbreviation $1/\eta_i = (a_i x_j + D^u_{ii}(x_i))$ and the convective field $\vec{\alpha}$. In contrast to \eqref{eq:drift:prediction:common} this expression depends now explicitly on the drift $\vec{f}$ as well as on the uncoupled diffusion matrix entries $D_{ii}^u(x_i)$.
\smallskip

\begin{figure}[h]
\hfill
\subfigure[]{\includegraphics[width=0.49\linewidth]{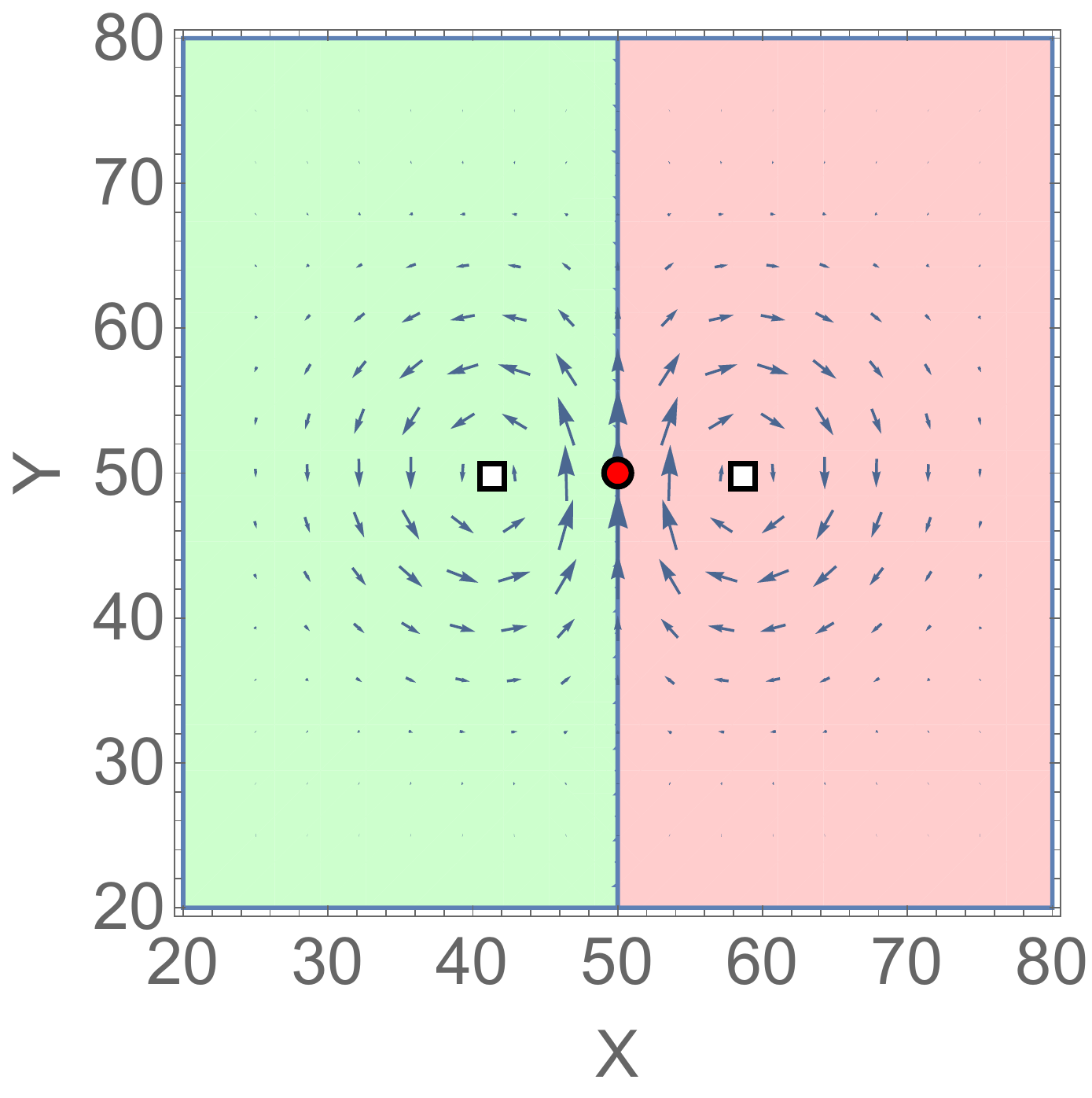}}
\hfill
\subfigure[]{\includegraphics[width=0.49\linewidth]{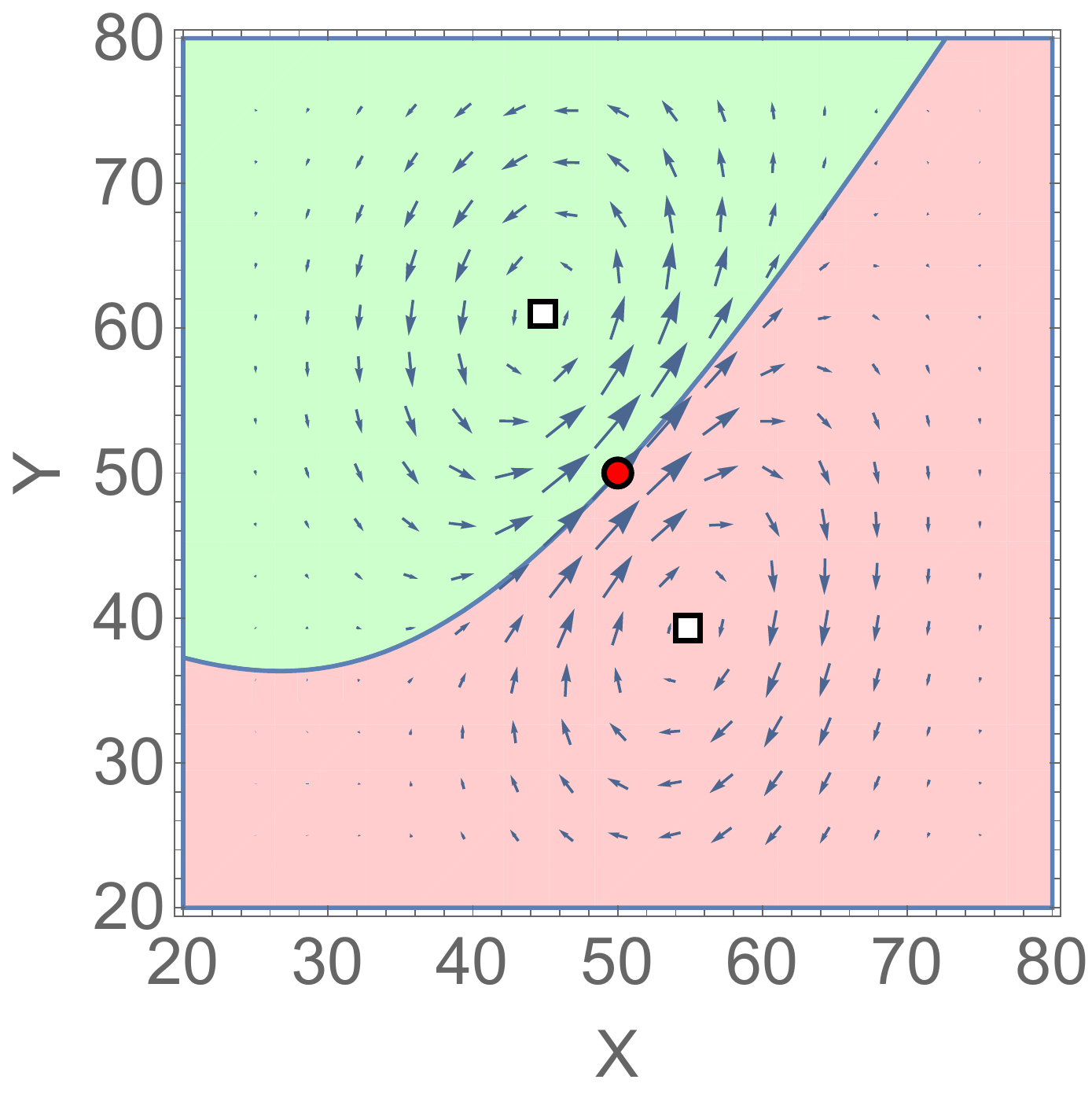}}
\hfill
\caption{
(Color online) Stationary probability current for the example system \eqref{eq:fp_example:drift} with linear coupling in the diagonal terms of the diffusion \eqref{eq:diff:diag:linear} with $D^u_{11}(x) = D^u_{22}(y) = d = 100$ and different coupling strengths:\\
(a)$a_1 = 1, a_2 = 0$\\
(b)$a_1 = 1, a_2 = 5$\\
 The green (lighter)/red (darker) background color indicates the positive/negative rotation sense predicted based on the sign of eq.~\eqref{eq:diff:lin:rotj}. The red dots indicate the position of the fixed point of $\vec{\alpha}$, the white rectangles the fixed points of $\vec{j}_s$.
\label{fig:j:diffusion}
}
\end{figure}

We can immediately see that this quantity is zero at the location of an $\vec{\alpha}$ fixed point. Therefore the leading order terms in $1/N$ vanish at this location, and we expect a dipole shape for the resulting $\vec{j}_s$ currents. To verify this prediction we have plotted the sign of expression \eqref{eq:diff:lin:rotj} together with the numerical solution of the corresponding Fokker-Planck equation in fig.~\ref{fig:j:diffusion}.

For this figure, we have used \eqref{eq:fp_example:drift} for the drift vector, as well as $\mathbf{D}^u = \begin{pmatrix}d&0\\0&d\end{pmatrix}$ for the underlying uncoupled diffusion matrix.

\smallskip

Due to the fact that all terms in \eqref{eq:diff:lin:rotj} are proportional to a component of $\vec{\alpha}$, dipole currents will emerge irrespective of the chosen drift vector, coupling or diffusion strengths of the system.

\subsubsection{More general conditions for dipole currents}

So far we have seen that the leading order term in $1/N$ vanishes (and thus higher order terms become important) if the quantity $\vec{\nabla}\times\mathbf{D}^{-1}{\vec{\alpha}}$ is zero at the location of an $\vec{\alpha}$ fixed point. It is straightforward to investigate for which types of reaction networks this condition holds. To this purpose, we write
\begin{align}
\begin{aligned}
    \vec{\nabla}\times\mathbf{D}^{-1}{\vec{\alpha}} =& 
    \left(\alpha_1\frac{\partial B_{12}}{\partial x} + B_{12} \frac{\partial \alpha_1}{\partial x} + B_{22} \frac{\partial \alpha_2}{\partial x} + \alpha_2 \frac{\partial B_{22}}{\partial x}\right)\\
    &- \left(\alpha_1\frac{\partial B_{11}}{\partial y} + B_{11} \frac{\partial \alpha_1}{\partial y} + B_{12} \frac{\partial \alpha_2}{\partial y} + \alpha_2 \frac{\partial B_{12}}{\partial y}\right)
\end{aligned}
\label{eq:barmagnetlike:rot1da}
\end{align}
with the abbreviation $B = D^{-1}$, i.e. $B_{12} = B_{21} = -\frac{D_{12}}{\Delta}$, $B_{11} = \frac{D_{22}}{\Delta}$, $B_{22} = \frac{D_{11}}{\Delta}$ with $\Delta = \text{det}(\mathbf{D}) = D_{11}D_{22} - D_{12}^2$.

\smallskip

At the location of a fixed point $\vec{x}_0$ of $\vec{\alpha}$, half of these terms vanish, and we can write
\begin{align}
    \begin{aligned}
    \vec{\nabla}\times\mathbf{D}^{-1}{\vec{\alpha}}(\vec{x}_0) =&
    B_{12} \left(\frac{\partial \alpha_1}{\partial x} - \frac{\partial \alpha_2}{\partial y}\right)\\
    &+ \left(B_{22} \frac{\partial \alpha_2}{\partial x} - B_{11} \frac{\partial \alpha_1}{\partial y}\right)
\end{aligned}
\label{eq:rotd1a:decomposition}
\end{align}

Whenever this quantity vanishes, the leading order term in $1/N$ is zero and higher-order terms need to be considered.

This is for instance the case when certain symmetries are present: Whenever the system includes an $x \leftrightarrow y$ - symmetry in $\vec{\alpha}$, the first term in \eqref{eq:rotd1a:decomposition} does not contribute. However this term will also be zero whenever the off-diagonal diffusion matrix elements $D_{12}$ vanish. Similarly, the second term will vanish if we choose the parameters of the system in a way that the symmetry condition $D_{11} \frac{\partial \alpha_2}{\partial x} = D_{22} \frac{\partial \alpha_1}{\partial y}$ is fulfilled at the location of an $\vec{\alpha}$ fixed point, but also if the drift vector is completely uncoupled and $\frac{\partial \alpha_2}{\partial x} = \frac{\partial \alpha_1}{\partial y} = 0$ holds. 

This means that in a system without any drift coupling and with vanishing off-diagonal diffusion elements, the leading order terms in $1/N$ will always be zero and the system will show dipole (or other higher order multipole) currents, irrespective of the chosen parameters. The system shown in fig.~\ref{fig:j:diffusion} clearly falls into this category.

\subsection{Stationary probability currents in
model reaction systems}
After having investigated how different coupling terms inside the drift or diffusion matrix affect the stationary probability currents $\vec{j}_s$, we now return to reaction systems. Any reaction that couples the two molecular species will introduce coupling terms at multiple locations inside the drift and diffusion terms simultaneously. For example the reaction

\begin{equation}
    X \overset{k}{\rightarrow} Y
\end{equation}
yields $-kx$ in $f_1(\vec{x})$, $+kx$ in $f_2(\vec{x})$ and $+kx$ in $D_{11}(\vec{x})$ and $D_{22}(\vec{x})$.

In the following, we will investigate on two generic example systems,
 one of which displays bistability, the other one a limit cycle.

\subsubsection{Positive feedback loop}
\label{sec:pos_feedback}
Positive feedback loops occur for instance in gene expression when a protein produced from one gene activates the expression of another gene and vice versa \cite{shoval2010snapshot}.

One possible implementation reads
\begin{align}
\begin{aligned}
 \emptyset &\overset{b_x }{\underset{d_x}{\rightleftarrows}} X\\
 \emptyset &\overset{b_y }{\underset{d_y}{\rightleftarrows}} Y\\
X &\overset{F_x(y)}{\rightarrow} \emptyset\\
Y &\overset{F_y(x)}{\rightarrow} \emptyset\\
\end{aligned}
\label{eq:pos:feedback}
\end{align}
with the Hill function 
\begin{equation}
F_y(x) = \frac{x^{n_x} \cdot m_y}{x^{n_x} + \theta_x ^{n_x}}.
\end{equation}

This leads to the drift vector
\begin{equation}
    \vec{f} = \begin{pmatrix}
    -d_x x + b_x N + \frac{m_y x^{n_y}N}{x^{n_y} + (\theta_y N)^{n_y}}\\
    -d_y y + b_y N + \frac{m_x x^{n_x}N}{x^{n_x} + (\theta_x N)^{n_x}}
    \end{pmatrix}
\end{equation}
and the diffusion matrix
\begin{equation}
    \mathbf{D} = \begin{pmatrix}
    d_x x + b_x N + \frac{m_y x^{n_y}N}{x^{n_y} + (\theta_y N)^{n_y}} \phantom{ABC} 0 \phantom{ABC}\\
    \phantom{ABC} 0 \phantom{ABC} d_y y + b_y N + \frac{m_x x^{n_x}N}{x^{n_x} + (\theta_x N)^{n_x}}\, .
    \end{pmatrix}
\end{equation}

\begin{figure}[H]
\hfill
\subfigure[]{\includegraphics[width=0.48\linewidth]{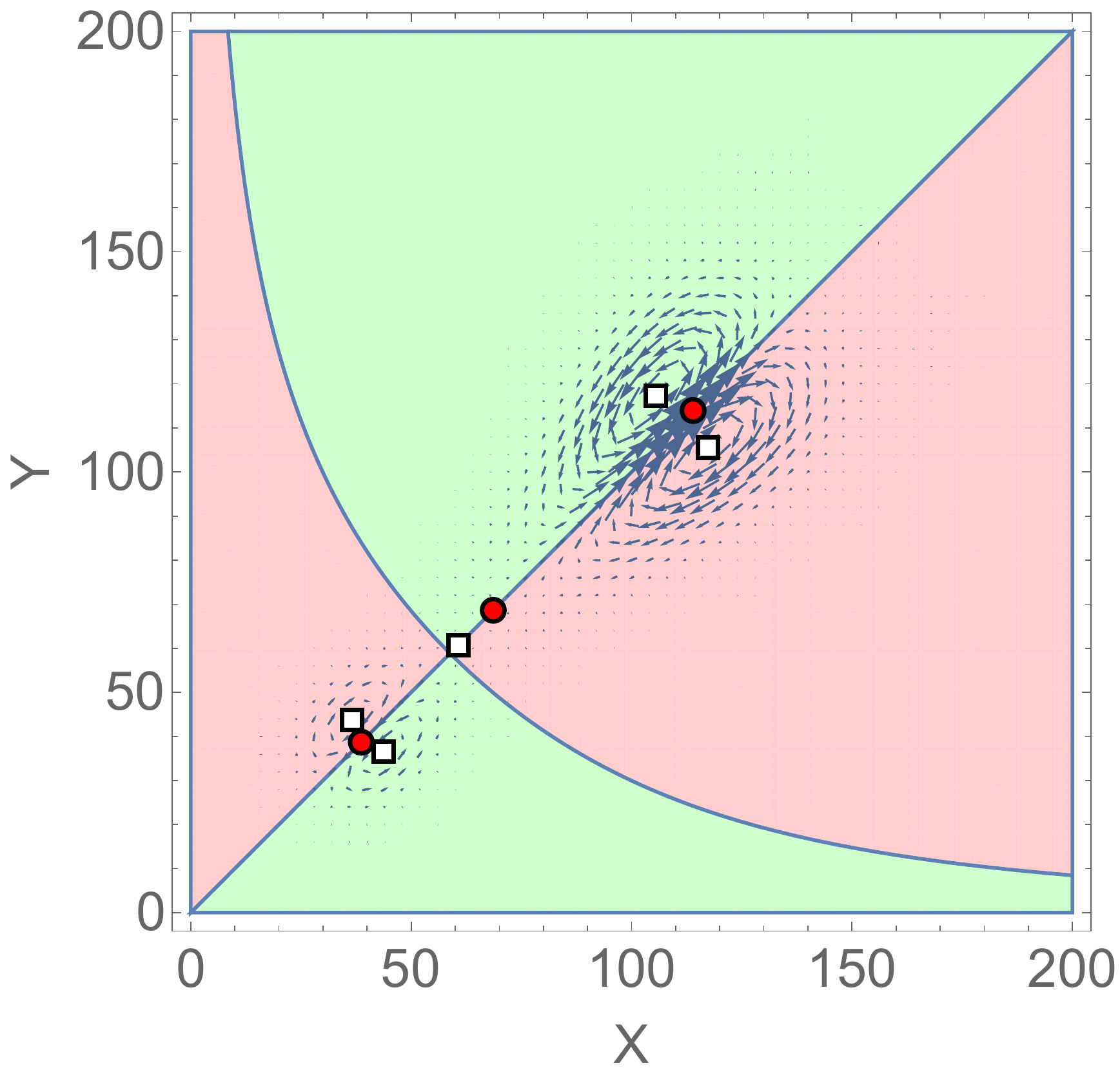}}
\hfill
\subfigure[]{\includegraphics[width=0.48\linewidth]{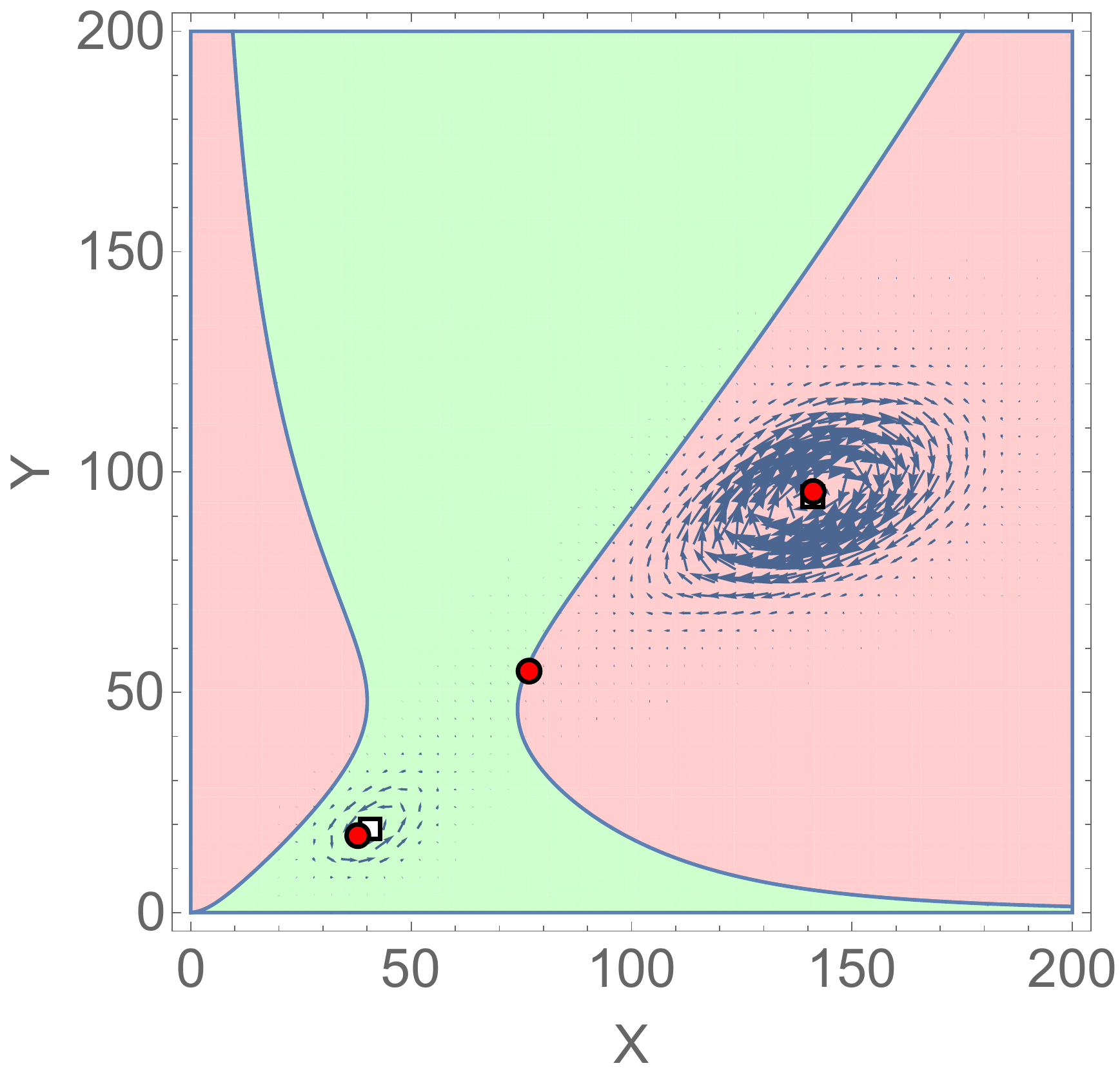}}
\hfill
\caption{\label{fig:feebackschleife}
(Color online) Stationary probability current (blue arrows) for system \eqref{eq:pos:feedback} with (a) symmetric and (b) asymmetric parameter values. The green (lighter)/red (darker) background color indicates the predicted positive/negative sense of rotation according to eq.~\eqref{eq:rot_sinn_j_allg}. The red dots indicate the position of the fixed points of $\vec{\alpha}$, the white rectangles the fixed points of $\vec{j}_s$. Parameter values are:\\
(a) $m_x = m_y = 1, n_x = n_y = 4, \Theta_x = \Theta_y = 4, d_x = d_y = 0.2, b_x = b_y = 0.34, N = 20$\\ (b) $m_x = m_y = 1, n_x = 4, n_y = 3, \Theta_x = \Theta_y = 4, d_x = 0.12, d_y = 0.22, b_x = 0.22, b_y = 0.15, N = 20$
}
\end{figure}

Depending on the parameters, this system can possess none, one or two stable $\vec{\alpha}$ fixed points in the positive quadrant. Fig.~\ref{fig:feebackschleife} shows two examples of a bistable system, with (a) symmetric and (b) asymmetric parameter choices.

In fig.~\ref{fig:feebackschleife} (a) $\vec{j}_s$ comprises essentially two opposing dipole fields.
A shape like this could be guessed based on the location of the $\vec{\alpha}$ fixed points and the sign of $\vec{\nabla}\times\mathbf{D}^{-1}\vec{\alpha}$.

Even though there is an $\vec{j}$ fixed point between the two dipoles, its location is displaced from the $\vec{\alpha}$ saddle point.

The asymmetric system in fig.~\ref{fig:feebackschleife} (b) on the other hand possesses two centers of opposite rotation sense in $\vec{j}_s$, which can also be guessed based on the fixed points of $\vec \alpha$ and on the sign of  $\vec{\nabla}\times\mathbf{D}^{-1}\vec{\alpha}$.
In this subfigure, the saddle point in $\vec{\alpha}$ has no visible counterpart in $\vec{j}_s$. 

However, since the shape of $\vec j_s$ must change continuously as the parameters are changed from Fig.~\ref{fig:feebackschleife}(a) to (b), the second pair of swirls must still be present, but has moved to a region where $p_s$ is very small, so that the numerical calculation cannot resolve it. This means that there is still a saddle point of $\vec j_s$, but it cannot be resolved either, and it does not coincide with the saddle of $\vec\alpha$. 

This is in contrast to section \ref{sec:leading_order_n}, where we have established that in leading order in $1/N$ $\vec{j}_s$ follows the height lines of $p_s$, which means that maxima, minima and saddle points of $p_s$ lead to the formation of $\vec{j}_s$ fixed points at the same locations. 
 Fig.~\ref{fig:feebackschleife}(b) implies that the expectation that is based purely on the leading order in $1/N$ is not correct, and that higher-order terms cannot be neglected for the parameter values used in this analysis. 

\begin{figure}[H]
\hfill
\subfigure[]{\includegraphics[width=0.48\linewidth]{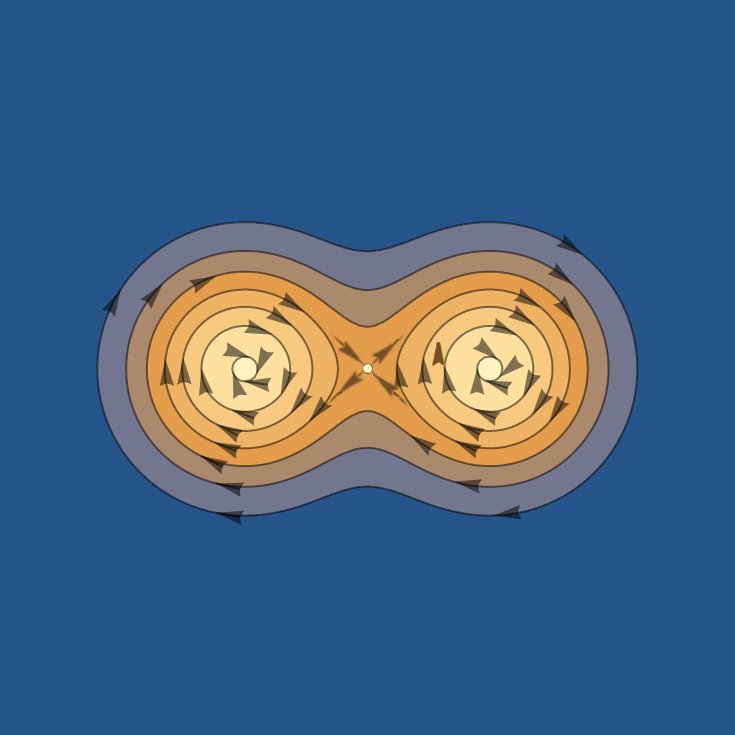}}
\hfill
\subfigure[]{\includegraphics[width=0.48\linewidth]{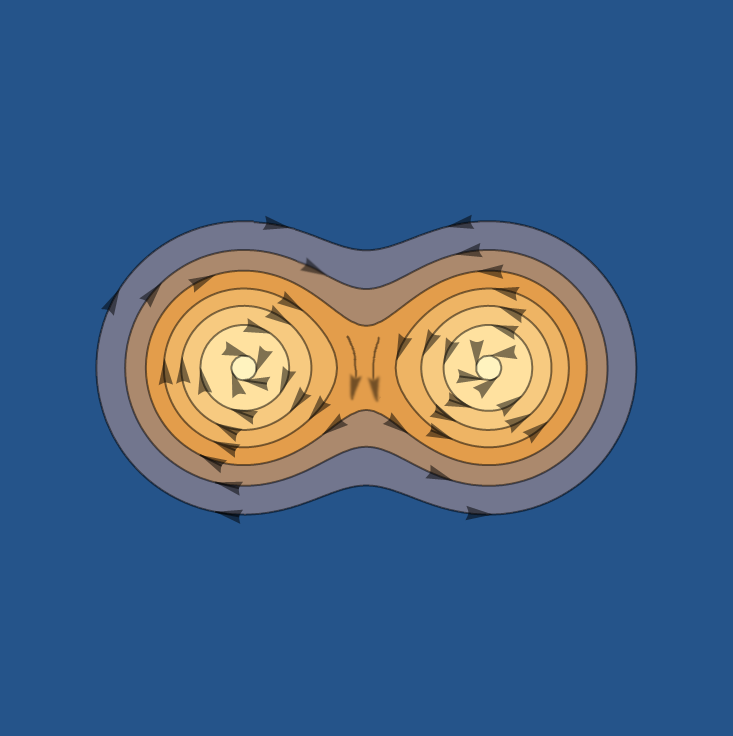}}
\hfill
\caption{\label{fig:p_saddle}\\
(Color online)\\
Color gradient: Height profile of $p_s$.\\
Arrows: direction of $\vec{j}_s$.\\
(a) A saddle point in $p_s$ between two maxima with equal rotation sense in $\vec{j}_s$ leads to the formation of a saddle in $\vec{j}_s$ at the location of the $p_s$-saddle.\\
(b) If a $p_s$-saddle lies between two maxima with opposing rotation sense in $\vec{j}_s$, the formation of a $\vec{j}_s$-saddle is prevented.
}
\end{figure}

We can find instances of saddles of $\vec\alpha$ without visible saddles of $\vec j_s$ also in a much broader context:  Depending on the rotation sense of $\vec j_s$ to the left and the right of a $p_s$ saddle, there are in general two options for the flow of $\vec{j}_s$ in the vicinity of a $p_s$-saddle. Both are shown in fig.~\ref{fig:p_saddle}.

In fig.~\ref{fig:p_saddle} (a), the $p_s$ saddle lies between two maxima with the same rotation sense of $\vec{j}_s$. In this case there is a matching $\vec{j}_s$ saddle. In fig.~\ref{fig:p_saddle} (b) however, the formation of a $\vec{j}_s$ saddle is impossible due the topology of the system: The two $\vec{j}_s$ centers with opposing rotation sense dictate that a constant $\vec{j}_s$ flow must be present between them, where the $p_s$-saddle lies. As this constant flow contradicts the calculated leading-order behavior, it must be caused by higher-order terms in $1/N$. 

\smallskip

\subsubsection{Rosenzweig-MacArthur model}
Our second example is a system that contains a limit cycle in its ODE description. We use the following variant of the Rosenzweig MacArthur model, written as a set of chemical reactions \cite{rosenzweig, mendler2018analysis}:

\begin{align}
\begin{aligned}
X & \overset{\beta}{\longrightarrow} 2 X\\
X + X &\overset{\delta}{\longrightarrow} X\\
X + Y &\overset{\frac{d}{1+Ax}}{\longrightarrow} 2 Y\\
X + Y &\overset{\frac{b-d}{1+Ax}}{\longrightarrow} Y\\
Y &\overset{c}{\longrightarrow} \emptyset\\
\emptyset &\overset{q}{\longrightarrow} X\\
\emptyset &\overset{q}{\longrightarrow} Y
\label{eq:rosenzweig:reactions}
\end{aligned}
\end{align}

Drift vector and diffusion matrix for this system read
\begin{equation}
\vec{f} = 
\begin{pmatrix}
q - R \cdot \frac{b}{d} + x \left(\beta - \delta (x-\frac{1}{N})\right) \\
q +R - c y
\label{eq:rosenzweig:drift}
\end{pmatrix}
\end{equation}
and
\begin{equation*}
\mathbf{D} = 
\begin{pmatrix}
q + R \cdot \frac{b}{d} + x(x-\frac{1}{N})\delta + x \beta & -R\\
-R & q + R + c y
\end{pmatrix}
\end{equation*}
with the abbreviation $R = \frac{d x y}{1 + A x}$.
\smallskip

In the following, we will focus on two different parameter sets: A system with a single, stable fixed point as well as a system with an unstable fixed point at enclosed by a limit cycle. Evaluating \eqref{eq:rot_sinn_j_allg} for this system yields a positive rotation sense of $\vec{j}_s$ around the respective fixed points for both parameter sets. Thus we expect a circular flow of $\vec{j}_s$ around these points. As   $\vec{\nabla}\times\mathbf{D}^{-1}\vec{\alpha}$ remains finite in the deterministic limit $N \to \infty$, we expect a good agreement between the fixed points of $\vec{\alpha}$ and the respective extrema of $p_s$. 

\begin{figure}[H]
\hfill
\subfigure[]{\includegraphics[width=0.48\linewidth]{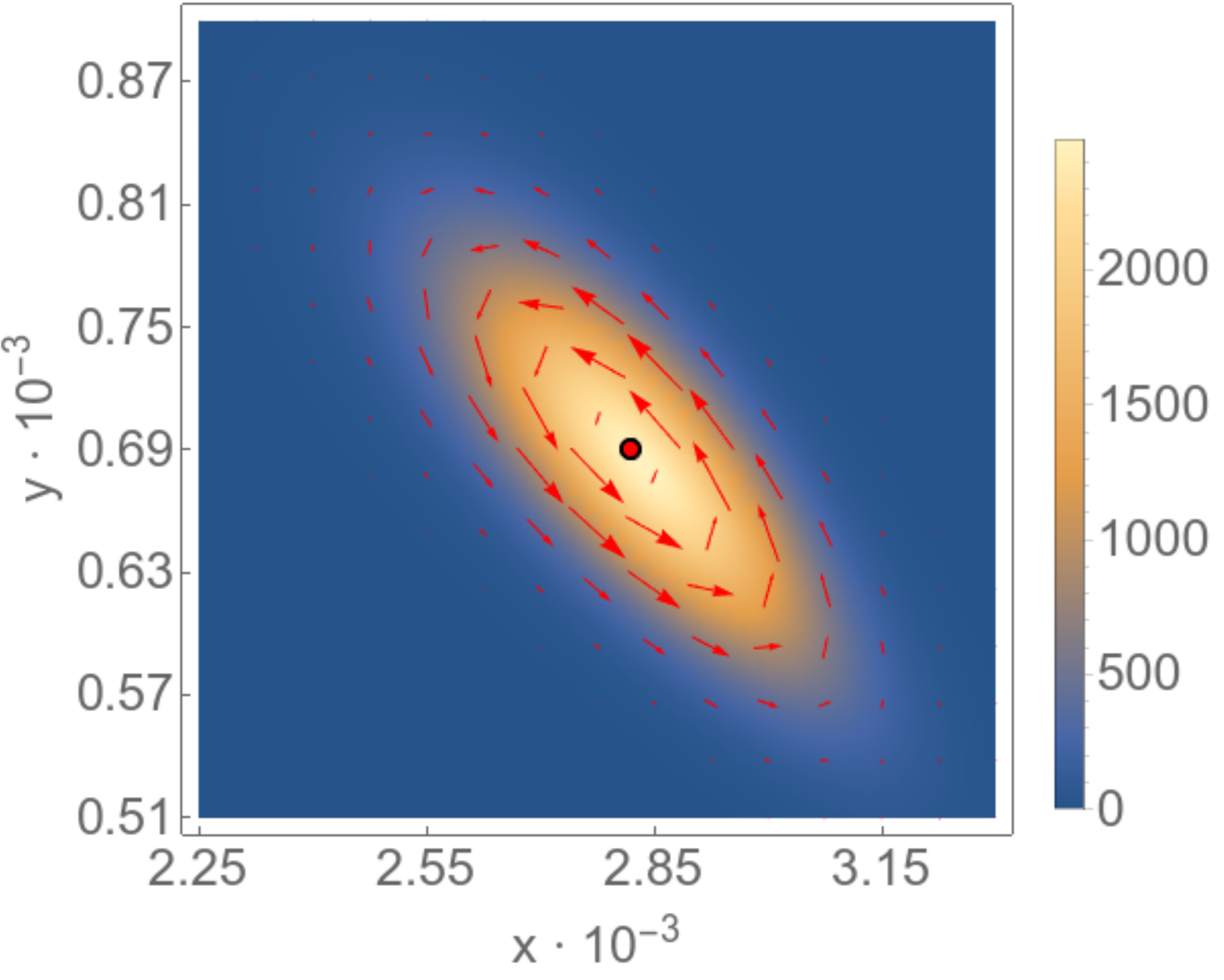}}
\hfill
\subfigure[]{\includegraphics[width=0.48\linewidth]{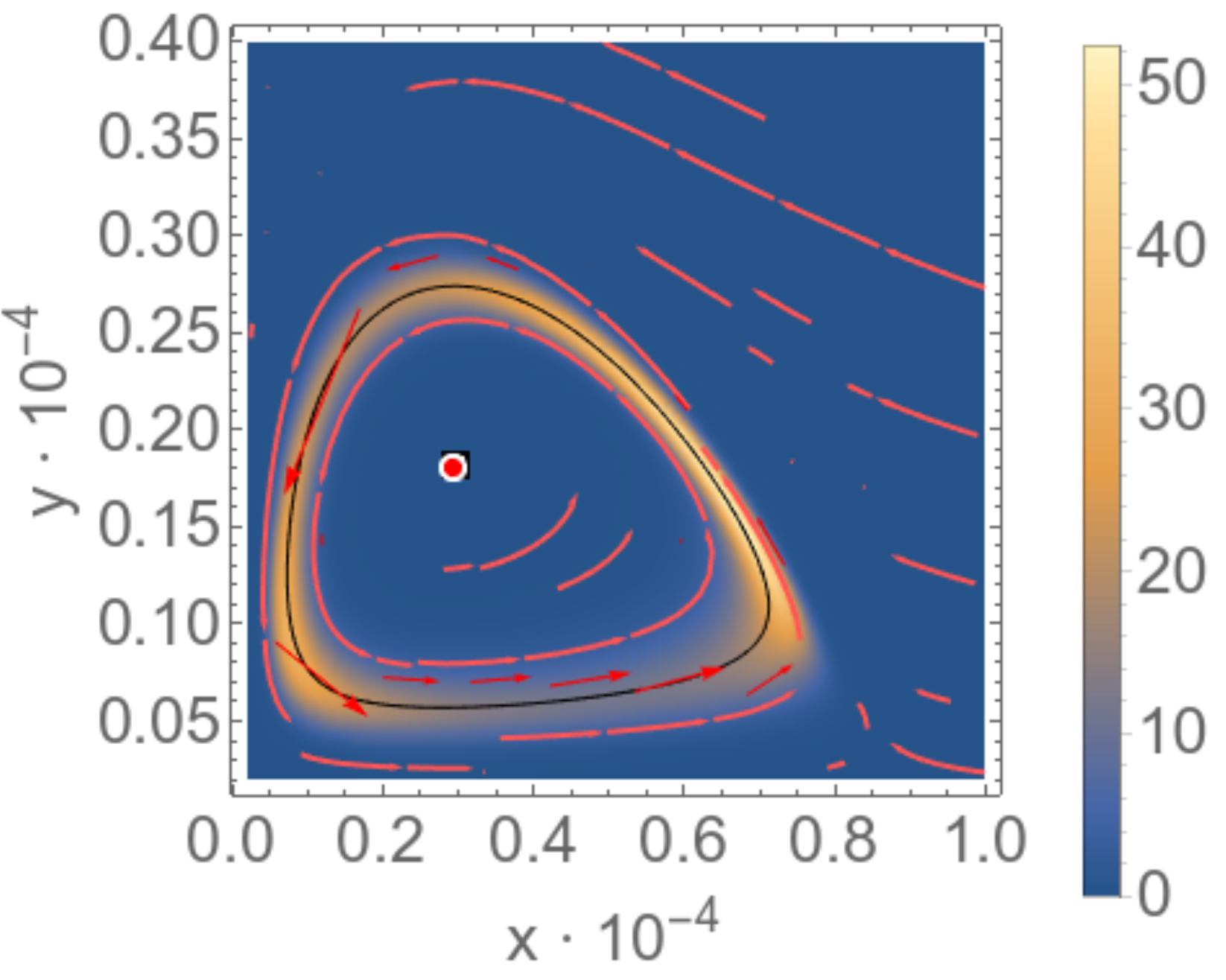}}
\hfill
\caption{\label{fig:rosenzweig}
(Color online) 
Stationary probability density (density plot in blue/yellow) and current (vector/streamline plot in red) for system \eqref{eq:rosenzweig:reactions} with parameter sets that (a) create a stable fixed point or (b) lead to an unstable fixed point enclosed by a limit cycle. The darker red vectors in fig.~(b) indicate the direction and strength of the vector field $\vec{j}$, whereas the lighter red streamlines indicate only its direction in places where its magnitude is too weak to use the former visualization. Furthermore the red dots in both panels indicate the fixed point in $\vec{\alpha}$, the white rectangle in (b) marks the minimum of the probability density and the black line in (b) shows the location of the limit cycle in $\vec{\alpha}$.
The parameters are:\\
(a)  $A = 2/3, \beta = 0.8, d = 0.65, c = 0.65, b = 1, \delta = 0.2, q = 0.01, \Omega = 1000$\\
(b) $A = 2/3, \beta = 0.9, d = 0.65, c = 0.65, b = 1, \delta = 0.1, q = 0.01, \Omega = 1000$
}
\end{figure}

The numerical solutions of the FPE for both parameter sets are shown in fig.~\ref{fig:rosenzweig}, where we can see that the extremum of $p_s$ agrees well with the fixed point of $\vec\alpha$ and of $\vec j_s$ in both cases.
Furthermore in panel (b) the probability density resembles a ridge that coincides with the location of the limit cycle in $\vec{\alpha}$. Along this ridge the probability density is non-constant, which leads to the formation of maxima and saddle points of $p_s$ along the ridge. 

As such maxima (saddle points) do not correspond to fixed points of $\vec \alpha$, but form when the probability flow along the ridge slows down (accelerates), we call them slow (fast) transient states. 

These findings are in contrast to section \ref{sec:leading_order_n}, where we have found that in leading order in $1/N$ $\vec{j}_s$ follows the height lines of $p_s$. In fact, the results of this section cannot be fully applied for limit cycles. The reason is that eq.~\eqref{eq:analytisch:p_solution} for the stationary probability distribution is not well defined in this case. This is because $\vec\alpha_\parallel$ points essentially in radial direction (when viewed from the minimum of the crater), resulting in a ring of fixed points, and concordantly in a ring of one-dimensional dynamical systems with no connection between them. The integral in the exponent in eq.~\eqref{eq:analytisch:p_solution} is therefore not well defined as it needs a separate integration constant for each radial direction. This means that $p_s$ cannot be specified based on 
$\vec\alpha_\parallel$ alone.

\smallskip

To see this explicitly, let us return to the rotationally symmetric model \eqref{eq:rotsymmmodel1} and consider the modification 
\begin{eqnarray}\label{eq:rotsymmmodel3}
    f_r(r) &=& ar - br^3  \nonumber\\
    f_{\varphi}(r, \varphi) &=& (\omega + \epsilon \cos(\varphi)) r
\end{eqnarray}
with positive $a,b,\omega,\epsilon$. For $\epsilon < \omega$, this system has a limit cycle at $r=\sqrt{a/b}$. Since $f_\varphi$ now depends on $\phi$, it is not rotationally symmetric any more. Nevertheless, the expression \eqref{eq:rotsymmmodel2} is still valid, and so is \eqref{eq:ps_rotsymm} for the stationary probability distribution. It contains an integral in the radial direction and leads to a maximum at $r=\sqrt{a/b}$. Based on $\vec \alpha_\parallel$ alone, different values of $\varphi$ are decoupled in the solution 
\eqref{eq:ps_rotsymm}.  Only by taking into account $\vec \alpha_\perp$ can we connect these different independent solutions, giving us the angular dependence of $A=A(\varphi)$. This can be done by starting from the stationarity condition \eqref{eq:divj_1N}. 
We write it in the following way, keeping only leading-order terms in each of the two components of $\vec\alpha$, i.e., order $N$ in terms with  $\vec{\alpha}_\parallel$ and order 1 in terms with $\vec{\alpha}_\perp $,
\begin{eqnarray}\label{eq:divj_1N2}
    0 &=& \mathbf{D} \vec{\alpha}_\parallel \cdot \vec{\nabla} p_s - \frac{1}{2 N} \vec{\nabla} (\mathbf{D} \vec{\nabla} p_s)\nonumber \\
    &+& (\vec{\nabla}\cdot \vec{\alpha}_\perp)  p_s + \vec{\alpha}_\perp \cdot \vec{\nabla}p_s .
\end{eqnarray}

The terms in the first line are of order $N$, those in the second line of order 1. If we want to find $\vec\alpha_\perp=\vec\alpha_\varphi$ along the ridge of the crater of $p_s$, we cannot neglect them, as we had done before, but we must keep them. In contrast to systems with an isolated fixed point, $p_s$ does not decay with a slope $\propto N$ in all directions, but it changes with a slope of the order of 1 in angular direction. This change can be found from evaluating the last line of \eqref{eq:divj_1N2}, giving 
 $A(\varphi) \propto 1/(\omega+\epsilon \cos(\varphi)$. This means that $p_s$ has a maximum at $\varphi = \pi$ and a saddle at $\varphi = 0$. 

With increasing $N$ the ridge of $p_s$ and the limit cycle of $\vec \alpha$ will agree better and better, as both must converge towards the behavior of the deterministic model. This means that Hopf bifurcations induced by changing parameters of the vector field $\vec f$ are reflected by transitions from mounds to craters in the stationary probability distribution when $N$ is large enough. In contrast, it was shown in \cite{becker2020relation} that Hopf bifurcations that are induced by a decrease of the system size $N$ are not reflected in such a transition. 

\subsection{Non-physical probability currents}
\label{sec:nonphysical_currents}
As a last set of examples, we will consider systems that show non-physical probability currents. Such currents emerge in 
systems the master equation of which shows detailed balance  \cite{ceccato2018remarks}, and they must be due to the approximations involved in deriving the FPE.
Our approach allows us to investigate the properties of these currents. 

As we have shown in section \ref{sec:int_cond}, stationary probability currents will emerge whenever $\vec{\nabla}\times\mathbf{D}^{-1}\vec{\alpha}~\neq~0$, as has also been noticed by Ceccato and Frezzato \cite{ceccato2018remarks}. 

Let us first consider the example reaction system investigated in \cite{ceccato2018remarks}:
\begin{align}
\begin{aligned}
\label{eq:ceccato:reactions}
    2 X &\underset{k_{-1}}{\overset{k_1}{\rightleftarrows}} X + Y\\
    X + Y &\underset{k_{-2}}{\overset{k_2}{\rightleftarrows}} Z\\
    Z &\underset{k_{-3}}{\overset{k_2}{\rightleftarrows}} 2 X
    \end{aligned}
\end{align}
As the quantity $n = n_X + n_Y + 2 n_Z$ is conserved, this system can be mapped onto a two-dimensional system which we can investigate with the methods described above.

\smallskip

Drift and diffusion for this system read:

\begin{align}
\begin{aligned}
    f_1 =& -(k_1 + 2 k_{-3}) x(x-1) + (k_{-1} - k_2)xy\\
    &+ (k_3 + \frac{1}{2} k_{-2}) \cdot (1 - x - y)\\
    f_2 =& k_1 x(x-1) + \frac{1}{2} k_{-2} (1 - x - y)\\
    &- (k_{-1} + k_2)xy
    \end{aligned}
\end{align}

\begin{align}
\begin{aligned}
    D_{11} =& (k_1 + 4 k_{-3}) x(x-1) + (2 k_3  + (k_{-1} + k_2)xy\\
    &+\frac{1}{2} k_{-2}) \cdot (1 - x - y)\\
    D_{12} =& k_1 x(x-1) + \frac{1}{2}k_{-2} (1-x-y) - k_2 x y - k_{-1} x y\\
    D_{22} =& k_1 x(x-1) + \frac{1}{2} k_{-2} (1 - x - y) + (k_{-1} + k_2)xy
    \end{aligned}
\end{align}

\begin{figure}[h]
\hfill
\subfigure[]{\includegraphics[width=0.48\linewidth]{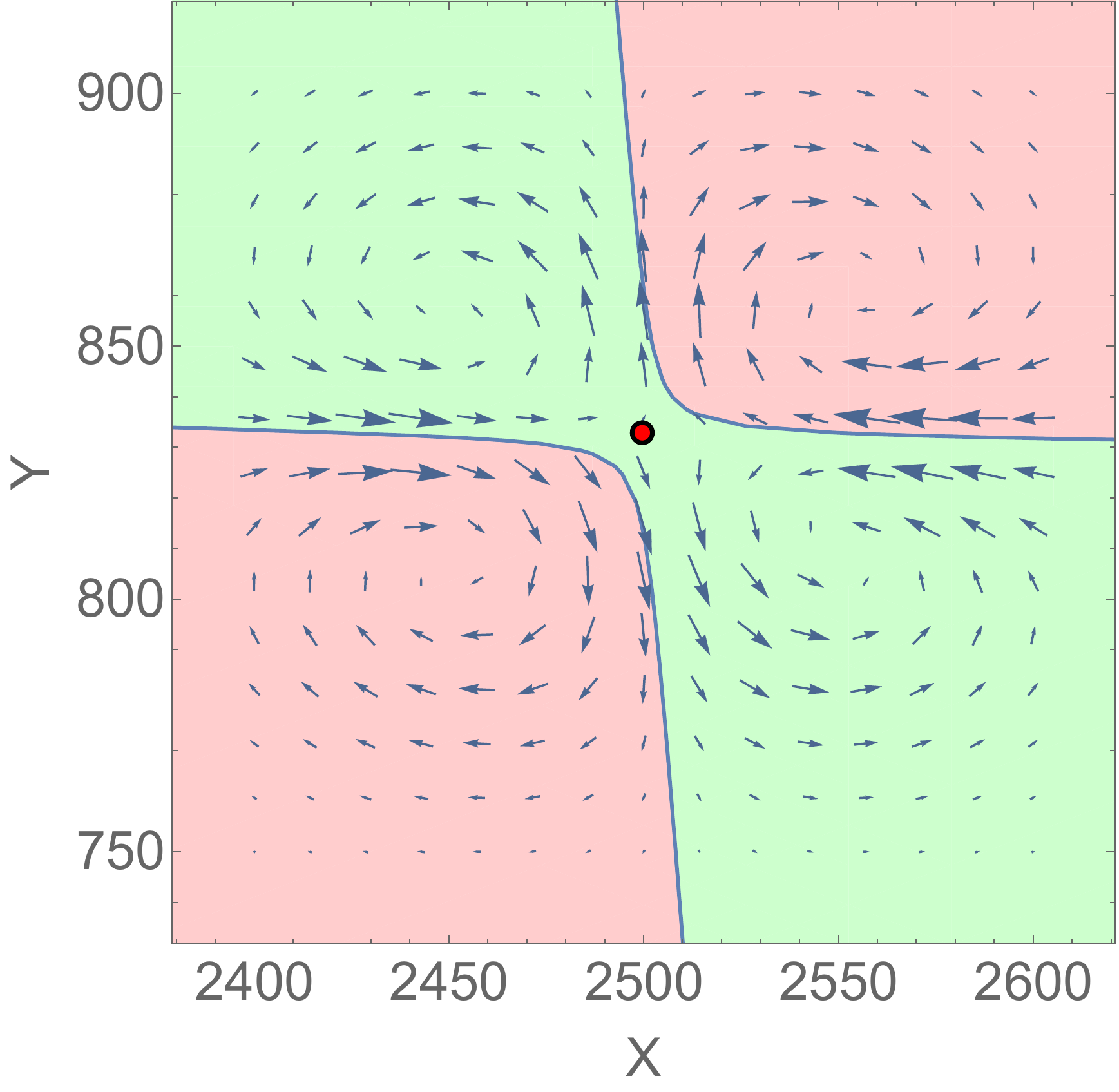}}
\hfill
\subfigure[]{\includegraphics[width=0.48\linewidth]{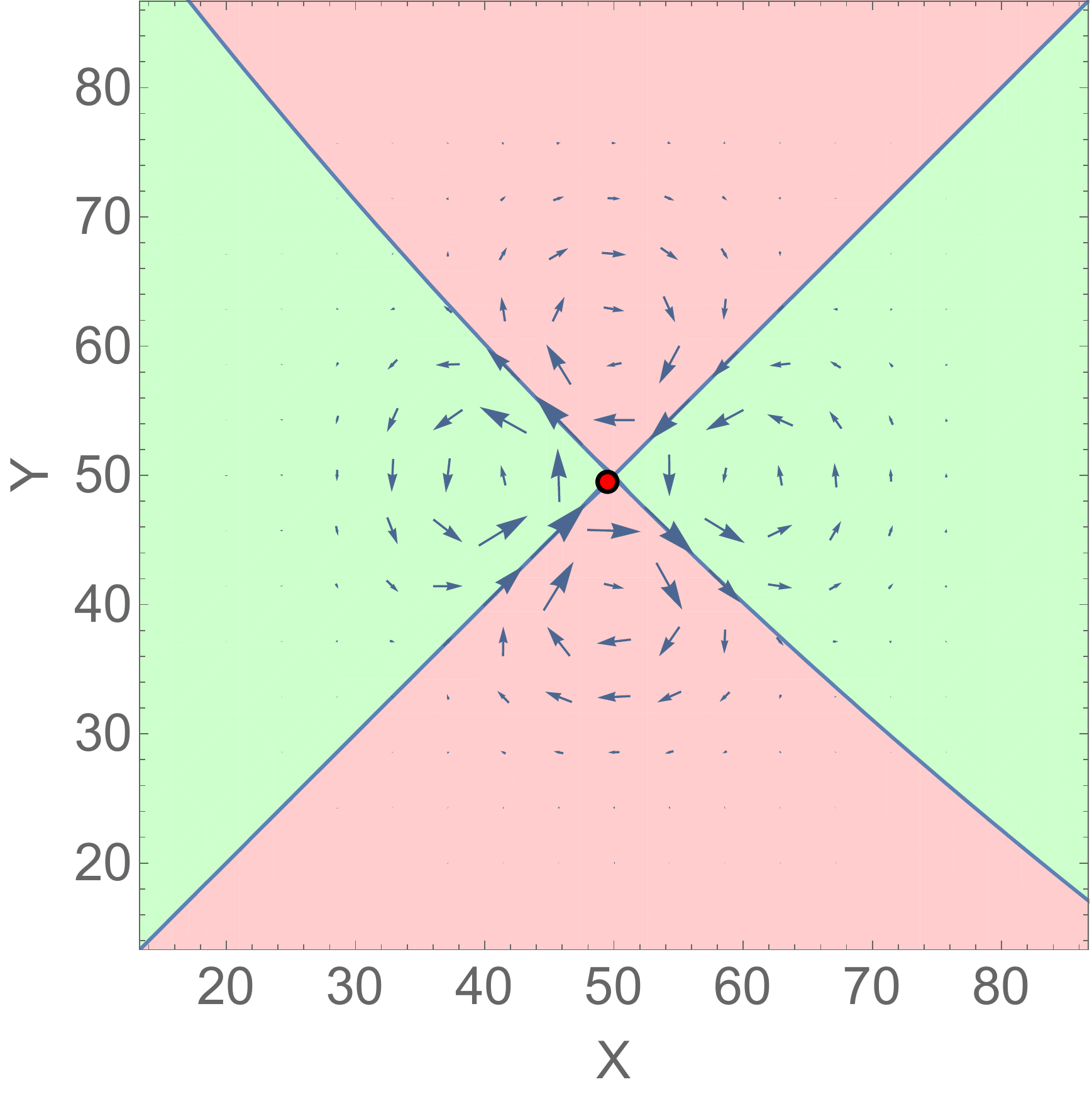}}
\hfill
\caption{
\label{fig:ceccato}
(Color online) Estimated rotation sense of $\vec{j}_s$ based on the sign of $\vec{\nabla}\times\mathbf{D}^{-1}\vec{\alpha}$ (background color; green (lighter): positive, red (darker): negative) and numerical results for $\vec{j}_s$ (blue vectors). The red dots indicate the fixed points of the convective field which coincide with the corresponding maxima of the probability density.\\
(a) The model \eqref{eq:ceccato:reactions} with parameters chosen as in \cite{ceccato2018remarks}, i.e. in our notation $k_1 = 1, k_{-1} = 3, k_2 = 2, k_{-2} = 500, k_3 = 75, k_{-3} = 0.1, n = 2\cdot 10^4$.\\
(b) The model \eqref{eq:nonphysical_system:reactions} with parameters $r = 50, s = \lambda = 1, N = 1$.
}
\end{figure}

The rotation sense of $\vec{j}_s$ as estimated based on $\vec{\nabla}\times\mathbf{D}^{-1}\vec{\alpha}$, as well as the numerical solution of the FPE are shown in fig.~\ref{fig:ceccato} (a).

One can see that the agreement on the topology of $\vec j_s$ between the simple prediction and the numerical result is pretty good.
The predicted rotation sense of $\vec{j}_s$ in \ref{fig:ceccato} can be interpreted as a condensed version of Fig. 3 in \cite{ceccato2018remarks} where only the sign of $\mathcal{R}(\tilde{\mathbf{\eta}})$ is shown.

\smallskip

Another example for non-physical currents in a system with detailed balance is given by 

\begin{equation}
    \vec{f} = \begin{pmatrix}
    (r - sx) + \lambda (y - x)\\
    (r - sy) + \lambda (x - y)
    \end{pmatrix}
\end{equation}
\begin{equation}
    \mathbf{D} = \begin{pmatrix}
    r + sx + \lambda (x+y) & -\lambda(x+y)\\
    -\lambda(x+y) & r + sy + \lambda (x+y)
    \end{pmatrix}
\end{equation}

which corresponds to the reaction system
\begin{align}
\begin{aligned}
    \emptyset &\underset{s}{\overset{r}{\rightleftarrows}} X\\
    \emptyset &\underset{s}{\overset{r}{\rightleftarrows}} Y\\
    X &\underset{\lambda}{\overset{\lambda}{\rightleftarrows}} Y
    \end{aligned}
    \label{eq:nonphysical_system:reactions}
\end{align}
that obeys detailed balance.

The non-physical stationary currents are shown in fig.~\ref{fig:ceccato} (b).
\smallskip

At the location of the fixed point, we have 
 $\vec{\nabla}\times\mathbf{D}^{-1}\vec{\alpha} = 0$. Interestingly,  the stationary probability current of this system does not take the shape of a dipole, as we found in previous examples with $\vec{\nabla}\times\mathbf{D}^{-1}\vec{\alpha} = 0$, but that of a quadrupole. Similarly to the dipole, this pattern satisfies the symmetry under exchanging $X$ and $Y$ in the reaction system, but in contrast to the dipole it has a saddle of $\vec j_s$ at the fixed point of $\vec \alpha$.

\smallskip
It is furthermore interesting to observe that this system satisfies $\vec{\nabla}\times\vec{\alpha} = 0$. The non-physical stationary probability currents are therefore due to the concentration dependence of the
diffusion matrix, which leads to $\vec{\nabla}\times\mathbf{D}^{-1}\vec{\alpha} \neq 0$.  Therefore one might call this kind of non-physical probability currents \textit{diffusion induced}. In contrast to this, the reaction system \eqref{eq:ceccato:reactions} shows already a preferred sense of rotation in its deterministic drift alone ($\vec{\nabla}\times\vec{f}\neq 0$), which is (given detailed balance in the master equation) sufficient for the occurrence of non-physical currents even if the corresponding diffusion matrix was a constant proportional to the identity matrix. Therefore one could call this type of non-physical probability currents \textit{drift induced}.

\section{Conclusion}

We have developed a method to estimate the topology of $\vec{j}_s$ from the convective field $\vec{\alpha}$ and the diffusion matrix $\mathbf{D}$ alone and without the need to solve the FPE numerically. Hereby we were able to prove that fixed points of $\vec{j}_s$ and $\vec{\alpha}$ and extrema of the stationary probability density $p_s$ coincide in leading order in the inverse system size $1/N$. 

This finding puts the suggestions made in \cite{mendler2018analysis} on a more solid foundation: In that paper, a technique to obtain phase portraits for 2-dimensional stochastic systems based on the Fokker-Planck equation was developed. However this technique relies without any proof on the assumption that $\vec{j}_s = 0$ at the location of an $\vec{\alpha}$ fixed point, so that fixed points of $\vec{\alpha}$ coincide with extrema of $p_s$. Even though this method works well for various examples, it was until now  unclear when this assumption breaks down. In the light of the results in this article, we can now specify that the stochastic phase portraits  suggested in  \cite{mendler2018analysis} are useful whenever the leading order terms in $1/N$ suffice to describe the system.

Furthermore we have seen that when the leading order term in $1/N$ is not sufficient, dipole currents in $\vec{j}_s$ can emerge, which means that $\vec{j}_s \neq 0$ at the location of an $\vec{\alpha}$ fixed point. Nevertheless, in the examples we have studied the distance between this $\vec{\alpha}$ fixed point and the corresponding $p_s$ extremum is still very small compared to the distance over which $p_s$ decays, which means that stochastic phase portraits would still be applicable even for theses cases. We were however not  able to identify the exact conditions under which this proximity occurs.

Together with the rotation sense of $\vec{j}_s$, which we estimated through the quantity $ \text{sgn}(\vec{\nabla}\times\mathbf{D}^{-1}\vec{\alpha}$), we were able to predict the topology of $\vec{j}_s$ for arbitrary reaction systems.

Using this method, we did an analysis of the different $j_s$ patterns that can emerge in chemical reaction networks with isolated coupling terms. The analysis of these simplified systems represents the first step towards understanding all the possible $j_s$ shapes that may emerge in more complex systems. Indeed we were able to identify a class of systems that only showed dipole-like currents given certain symmetric parameter choices as well as a class of systems that showed higher-order currents irrespective of the parameters of the system. 

Throughout these investigations we saw that our simple estimation formula for the rotation sense of $\vec{j}_s$, which is technically only correct for systems with constant, isotropic diffusion or for very large system sizes, gave quite accurate results for all the examples investigated. 

This work has thus laid a foundation on which further explorations of stationary currents can build.

\section*{Acknowledgments}
We thank J. Falk and L. Becker for helpful discussions. 
We acknowledge support by the German Research Foundation and the Open Access Publishing Fund of Technische Universität Darmstadt.

\bibliography{references}

\end{document}